\documentclass[3p]{elsarticle}
\usepackage{amsmath}
\usepackage{color}
\usepackage{amssymb}
\journal{Physica A}
\begin{document}
\bibliographystyle{elsarticle-num}
\begin{frontmatter}

\title{Thermodynamical behavior of the Blume-Capel model in the vicinity of its tricritical point}

\author[ifepe]{M\'ario  J. G. Rocha-Neto\corref{cor1}}
\ead{mario.neto@garanhuns.ifpe.edu.br}
\address[ifepe]{Instituto Federal de Educa\c c\~ao Ci\^encia e Tecnologia de Pernambuco, 55299-390, Garanhuns, Brazil}
\cortext[cor1]{Corresponding author} 

\author[caufpe]{G. Camelo-Neto}
\ead{gustavo.camelont@ufpe.br}
\address[caufpe]{N\'ucleo Interdisciplinar de Ci\^encias Exatas e da Natureza, Universidade Federal de Pernambuco, \\55.014-900, Caruaru, Brazil.}

\author[dfufpb]{E. Nogueira Jr.}
\ead{enogue@fisica.ufpb.br}
\address[dfufpb]{Departamento de F\'isica, Universidade Federal da Para\'iba, 58051-970, Jo\~ao Pessoa, Brazil.}

\author[dfufpe]{S. Coutinho}
\ead{sergio@ufpe.br}
\address[dfufpe]{Laborat\'orio de F\'{\i}sica Te\'orica e Computacional,
Departamento de F\'{\i}sica, Universidade Federal de Pernambuco, 50670-901, Recife, Brazil}

\date{\today}
%%%%%%%%%%%%%%%%%%%%%%%%%%%%
\begin{abstract}
We investigate the thermodynamic properties of the zero-field Blume-Capel model in the vicinity of its tricritical point (TCP). We calculate the internal energy, entropy, magnetization, quadrupole moment densities, and their response functions (specific heat and susceptibility) by employing an exact numerical recursion procedure for the model defined in a hierarchical lattice of fractal dimension $d$. We explore the scaling behavior of the isothermal and constant crystal-field-specific heat as a function of the temperature and the reduced crystal-field parameter along the ferromagnetic and the \emph{ordered paramagnetic} phase frontier. Results achieved for systems with dimensions $d=2$ and $3$ exhibit the main features of the continuous and first-order transitions in the TCP neighborhoods. We also probe the phase coexistence in the $\lambda$ diagram and the latent heat in the neighborhood of the tricritical points locus.
\end{abstract}

%%%%%%%%%%%%%%%%%%%%%%%%%%%


\end{frontmatter}

\section*{Highlights}
\begin{itemize}
\item The thermodynamics of the Blume-Capel model is studied near its tricritical point.
\item The exact local quadrupolar density, internal energy, and entropy are calculated.
\item  The quadrupolar susceptibility and specific heat are studied by finite-size scaling.
\item Quadrupolar susceptibility and specific heat exhibit a clear first-order transition.
\item The $\lambda$ diagram, entropy, and latent heat are probed near the TCP locus.
\end{itemize}
%%%%%%%%%%%%%%%%%%%%%%%%%%%%%
\section{Introduction}\label{sec1}
The spin-one Ising model, known as the Blume-Capel (BC) model, is one of the most well-studied spin models since its independent formulation by Capel \cite{capel66} and Blume \cite{blume66}, the latter aiming to describe the first-order magnetic phase transition observed in UO$_2$. Despite its simplicity, the BC phase diagram exhibits a second-order transition, a first-order transition, and a tricritical point (TCP). {\em Tricritical point}, a name coined by Griffiths \cite{griffiths70}, denotes the simplest multicritical point resulting from the confluence of three critical lines. BC is the simplest model to display a TCP in the three-field parameter space $(T,D,H)$, where $T$ is the temperature, $D$ is a nonordering field generated by the anisotropic crystal field and $H$ is an external ordering field, respectively. The BC model at $H = 0$ presents in the $(T,D)$ manifold a TCP as the point where the continuous transition critical line ends in a first order like one. 

Theoretical studies of the BC model have been and are being carried out using several techniques, such as those mentioned by us in the reference \cite{rochaneto18}. In particular, its tricritical behavior was earlier investigated in detail by the series method \cite{saul74}. The locus of the tricritical point in the phase diagram $(T,D)$ depends greatly on the details of the model and the methodology used to investigate the phase transition. For example, the TCP locus of the emblematic square-lattice BC model has been searched by accurate Monte Carlo computational simulation and transfer matrix methods \cite{kwak15,jung17,zierenberg17,blote19,fytas19,azhari22}. In particular, the recent review by Zierenberg et al. \cite{zierenberg17} gives several close values of its location obtained by such methodologies. 

The existence of TCP in the generalized BC model with integer $S>1$ or half-odd spin variable \cite{butera18,lafhal19,azhari20}, and with random field or competing exchange interactions \cite{albayrak13,albayrak17,santos18,guerrero19,karimou20,sumedha20} was also recently explored. For instance, reference \cite{butera18} investigate in detail by series expansions the cases $S=1$ and $S=3/2$ in the $2d$-square and $3d$-cubic lattices (\emph{sc} and \emph{bcc}) where TCP was found for the $S=1$ but not for the $S=3/2$ model, confirming the general indication emerging from the mean field approximation (MF). Moreover, in past \cite{kaneyoshi87,bobak97,buendia97} and recent years \cite{azhari20,oitmaa06,selke10,madani15,karimou17,yessoufou18,albayrak19,zahir21}, there was significant theoretical and experimental interest in investigating the phase diagram of \emph{mixed} Ising models, with semi-integer and integer $S$ spins located in interpenetrated lattices and subjected to a quadratic single-ion crystal field acting on integer $S$ spins. In the case of the mixed model composed with spins $1/2$ and $S$ ($S >1/2$) the absence of TCP in the square lattice was demonstrated through an exact procedure and confirmed by Monte Carlo simulation \cite{buendia97,dakhama18}, although TCP has been reported in these models using approximate methods (MFT) in systems with $d=2$ \cite{kaneyoshi91,jurcisinova22}. However, in three-dimensional systems, TCP was observed by Monte Carlo simulation in such mixed spin models on cubic lattices \cite{bobak97,selke10}. It is also important to point out here that for models with \emph{integer} mixed spins ($S=1$ and $S=2$), the TCP was found in the phase diagram of the model defined in $2d$ and $3d$ hypercubic interpenetrated lattices by mean field methods (MFT) or Monte Carlo simulation \cite{wei06}.

Pure, diluted, or disordered spin models (Ising, Potts, XY, Heisenberg, etc.) defined in hierarchical lattices have been used recently as a framework to study different systems in distinct contexts \cite{artun20,devre22,gurleyen22,tunca22,akin22,artun23}. The scale-invariant feature, which is the trademark of these lattices allows for the obtainment of analytical and exact solutions and hence the analysis of the intrinsic physical properties of the models. In the particular case of the family of diamond hierarchical lattices with fractal dimension $d$, the exact solutions found correspond to those obtained in the Migdal-Kadanoff approximation \cite{migdal75,kadanoff76,griffiths82} of the renormalization group in real space for the case of hyper-cubic lattices with the same $d$ dimension.

This work investigates the thermodynamic behavior of the Blume-Capel model defined on the family of diamond hierarchical lattices (hereafter DHL) of fractal dimensions $d$.  Internal energy, entropy, magnetization, quadrupole moment densities, and their response functions (specific heat and susceptibility) are accurately calculated and their properties are explored close to the critical transition lines, in particular, in the vicinity of the observed tricritical point. These quantities are exact and numerically obtained from the local magnetization and quadrupole moment densities, and from the internal energy by summing its exact local values generated by the Migdal-Kadanoff renormalization group combined with a recursive procedure presented and discussed in a previous paper by the authors \cite{rochaneto18}.  The response functions, the specific heat, and the susceptibility are extracted from the numerical derivatives of the internal energy and the quadrupole moment density, and finally, the entropy and the enthalpy are reconstructed by numerical integration. The phase diagram $(T \times D)$ of the BC model shows a \emph{reentrant} aspect, as discussed by the authors in \cite{rochaneto18}. As far as the temperature is lowered from higher values, within a certain interval of the values above the critical crystal field, the system undergoes a continuous phase transition from the \emph{disordered} paramagnetic phase (P) to the ordered ferromagnetic phase (F) followed by an unexpected phase transition to an \emph{ordered} paramagnetic phase (OP), which is characterized by zero magnetization but with majority of spins in the state $S = 0$. The F-OP frontier of the latter transition contains the locus of TCP, the transition being continuous (discontinuous) above (below) TCP as explicitly displayed in Figure \ref{Fig2}. This paper specifically focuses on the behavior of the quantities mentioned above in the \emph{reentrant} region of the phase diagram, in particular, in the vicinity of TCP. The reentrant phenomenon in a phase transition, which occurs in many different systems, has a long and rich history \cite{cladis88} since its probable first experimental observation in the ferroelectric transition at \emph{Rochelle salt} \cite{valasek1921}. The occurrence of reentrant boundaries in phase diagrams is related to the energy-entropic imbalance between phase configurations in the vicinity of the ground state at very low temperatures. In some systems, this imbalance results from two ingredients: the degeneracy of the ground state due to the existence of random conflicting interactions, which cause frustrated states \cite{cladis88,kitatani86,santos87}, and/or the inhomogeneity in the lattice site coordination number distribution, as shown in decorated spin models in regular lattices\cite{azhari22,santos87}. For the BC model defined on DHLs both ingredients contribute to the reentrant transition that occurs at low temperatures and values of the crystal field parameter just above its critical value at $T = 0$. In reference \cite{rochaneto18}, we showed that such transitions occur with finite discontinuities in the order parameters in this region of the phase diagram.   In the present work, the characterization of the nature of these phase transitions and the location of the tricritical point on the boundary between the F and OP phases are analyzed here through the behavior of the above-mentioned thermodynamic quantities.

This paper is organized as follows. In Section \ref{sec2}, we present the description of the model and its phase diagram, in addition to a brief description of the methodology developed in detail in \cite{rochaneto18}. Section \ref{sec3} is dedicated to the numerical calculation of the density of quadrupole moments and to exploring its isothermal susceptibility. In the following Section \ref{sec4} we calculate de exchange-internal energy and pay attention to the constant-field and isothermal specific heats. In Section \ref{sec5} the entropy and the enthalpy (latent heat) are numerically obtained by integration. Finally, in Section \ref{sec6} we summarize all results and conclusions.

\section{Model and Methodology}\label{sec2}
\subsection{The model Hamiltonian}
The Hamiltonian $\mathcal{H}_{\textsf{BC}}$ for the Blume-Capel model can be written as
\begin{equation}\label{eq1}
\mathcal{H}_{\textsf{BC}}=-K \sum_{\langle ij\rangle}  S_iS_j+ D \sum_{i} S_i^2 - H \sum_i S_i
\end{equation}
where $K$ ($K>0$) is the exchange coupling constant interaction associated with the nearest neighbors $\langle ij \rangle$ pair of spins, $D$ labels the local crystal-field strength and $H$ is an external magnetic field acting on the site spin variables $S_i$, which take  the values $S_i=0,\pm 1$. The crystal-field coupling $D$ governs the fraction of $S_i=0$ spins. 

 The total average energy $\langle \mathcal{H}_{\textsf{BC}}\rangle$, which can be formally defined by $\mathcal F_N+ T \partial \mathcal F_N/\partial T$, $\mathcal F_N$ being the Helmholtz free-energy $\mathcal F_N (T,H,D) = - k_B T \ln [\textrm{Tr}_{\{S_i\}} \exp(-(k_B T)^{-1} \mathcal{H}_{\textsf{BC}})] $, results
\begin{equation}
\label{eq2}
 \langle \mathcal{H}_{\textsf{BC}}\rangle =-K \sum_{\langle ij\rangle}  \langle S_iS_j \rangle + D \sum_{i}\langle  S_i^2 \rangle-H \sum_i \langle S_i\rangle = -U_K+U_{D}-U_H \ , 
\end{equation}
where $\langle \ldots \rangle$ denotes the canonical ensemble average in the thermodynamic limit.  $\langle \mathcal{H}_{\textsf{BC}}\rangle$ can be partitioned in  the ``internal" exchange energy contribution from the bond spins $U_K$, in addition to the energy contributions $U_{D}=DQ$  and $U_{H}=HM$, arising from the interaction of the spin system with the ``external'', crystal field $D$ and magnetic field $H$, respectively,  $M$ being the magnetization and $Q$, the quadrupolar moment.

This spin-one model has two densities, the magnetization density $m = M/N_s$ and the quadrupole moment density $q=Q/N_s$, given respectively by
\begin{equation}\label{eq3}
m= \frac{1}{N_s}\sum_i \langle S_i\rangle, \qquad \textsf{and} \qquad q= \frac{1}{N_s}\sum_i \langle S_i^2\rangle,
\end{equation}
where $N_s$ is the total number of  lattice sites.

\begin{figure}[ht]
\begin{center}
\includegraphics[width=10cm]{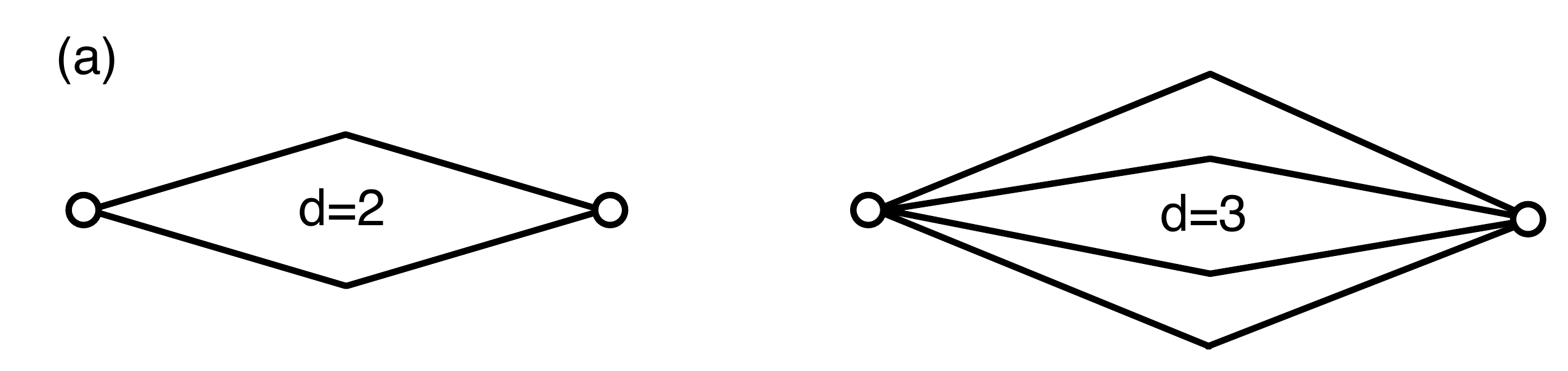}\\[2ex]
\includegraphics[width=14cm]{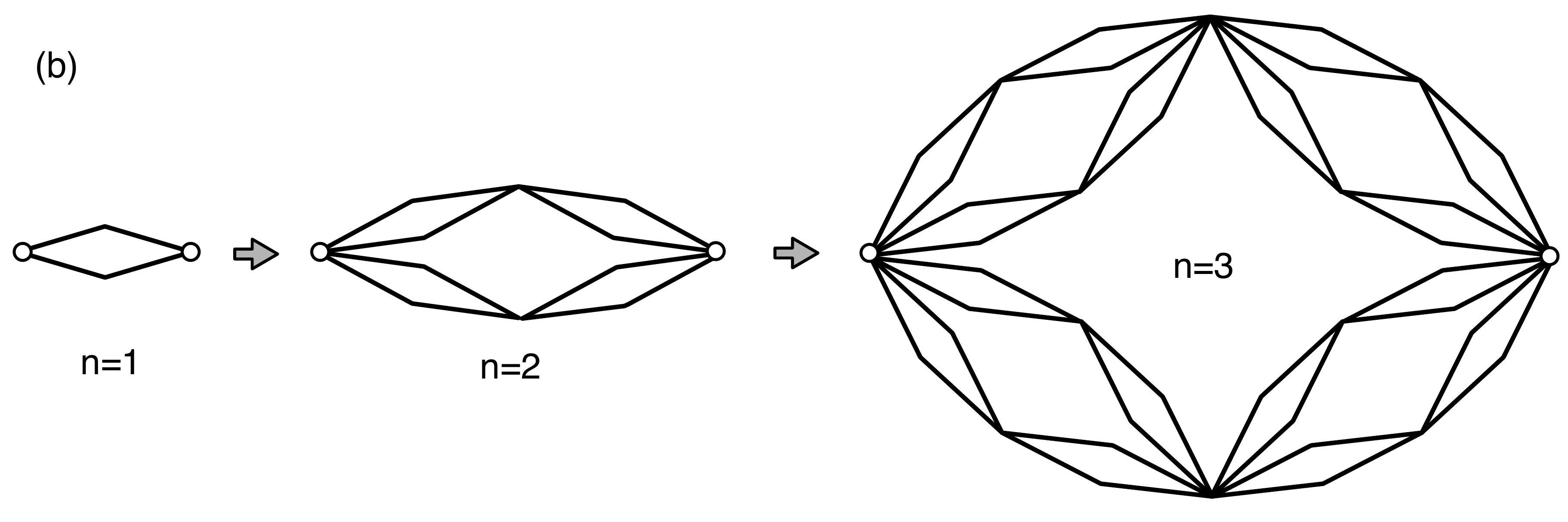}
\caption{Diamond hierarchical lattice with scaling factor $2$: (a) Basic units for the DHL with dimensions $d=2$ and $d=3$; (b) first three generations of the \emph{inflation} process to construct the DHL with dimension $d=2$. The gray arrow indicates the inflation process. Open circles denote the root sites.}
\label{Fig1}
\end{center}
\end{figure}

From equations (\ref{eq2}) and (\ref{eq3}) we construct the temperature- and field-dependent reduced energy  density $\varepsilon_{[n]}$ (in units of $K$) for a finite DHL with $n$ generations as
\begin{equation}
\label{eq4}
\varepsilon_{[n]} = - \frac{1}{N_{b}^{(n)}}\sum_{\langle ij\rangle}  \langle S_iS_j \rangle +  \frac{\alpha}{N_{s}^{(n)}} \sum_{i}\langle  S_i^2 \rangle-\frac{h}{N_{s}^{(n)}} \sum_{i}\langle  S_i \rangle = - u +  u_\alpha - u_h
\end{equation}
 where $u_\alpha=\alpha q$ and $u_h= h m$,  with $\alpha= D/K$ and $h=H/K$  being the reduced  crystal-field and external magnetic field coupling parameters, respectively.    $N_{b}^{(n)}=2^{d\,n}$ and $N_{s}^{(n)}=2+2^{d-1}[2^{d\,n}-1]/(2^d -1)$ are respectively the number of bonds and sites of a DHL with $n$ generations with scaling factor $2$ and fractal dimension $d$,  which basic units for $d=2$ and 3 are sketched in Figure \ref{Fig1}(a).  Figure \ref{Fig1}(b) illustrates the inflation process to construct the $d=2$ DHL up to the $n=3$ generation.

In this work, we consider the BC model defined in the diamond hierarchical lattice (DHL) with dimensions $d=2$ and $d=3$ and scaling factor 2. References \cite{griffiths82,kaufman84}  provide a broad discussion of the properties of spin models defined in these lattices.   The generalization of the present study to consider the BC model defined on DHL's of fractal or higher integer dimensions is straightforward,  but laborious.

\subsection{The real space renormalization process and phase diagram}

The real-space renormalization group equations for the temperature reduced coupling constants $J= \beta K$ and $\Delta=\beta D$ of the Blume-Capel Hamiltonian defined by equation (\ref{eq1}) at zero magnetic field, where $\beta=1/k_B T$, $k_B$ being the Boltzmann constant and $T$ the absolute temperature, result as
\begin{align}
\label{eq5}
   J'&=\ln \left[\frac{1+ 2 e^{-\Delta} \cosh(2 J)}{1+ 2 e^{-\Delta}}\right]^{p/2} \mbox{,}    \\
   \Delta' &= \Delta - \ln \left [\frac{1+ 2 e^{-\Delta} \cosh( J)}{1+ 2 e^{-\Delta}}\right]^p \mbox{.}  \label{eq6}
\end{align}
The prime superscripts on the above equations indicate de corresponding renormalized quantities after the decimation of the internal sites of the basic unit and $p$ is the number of connections between the root sites of the basic unit, which is related to the lattice dimensionality $d$ by $d=1+\ln p /\ln 2$.\\

\begin{figure}[ht]
\begin{center}
\includegraphics[width=8cm]{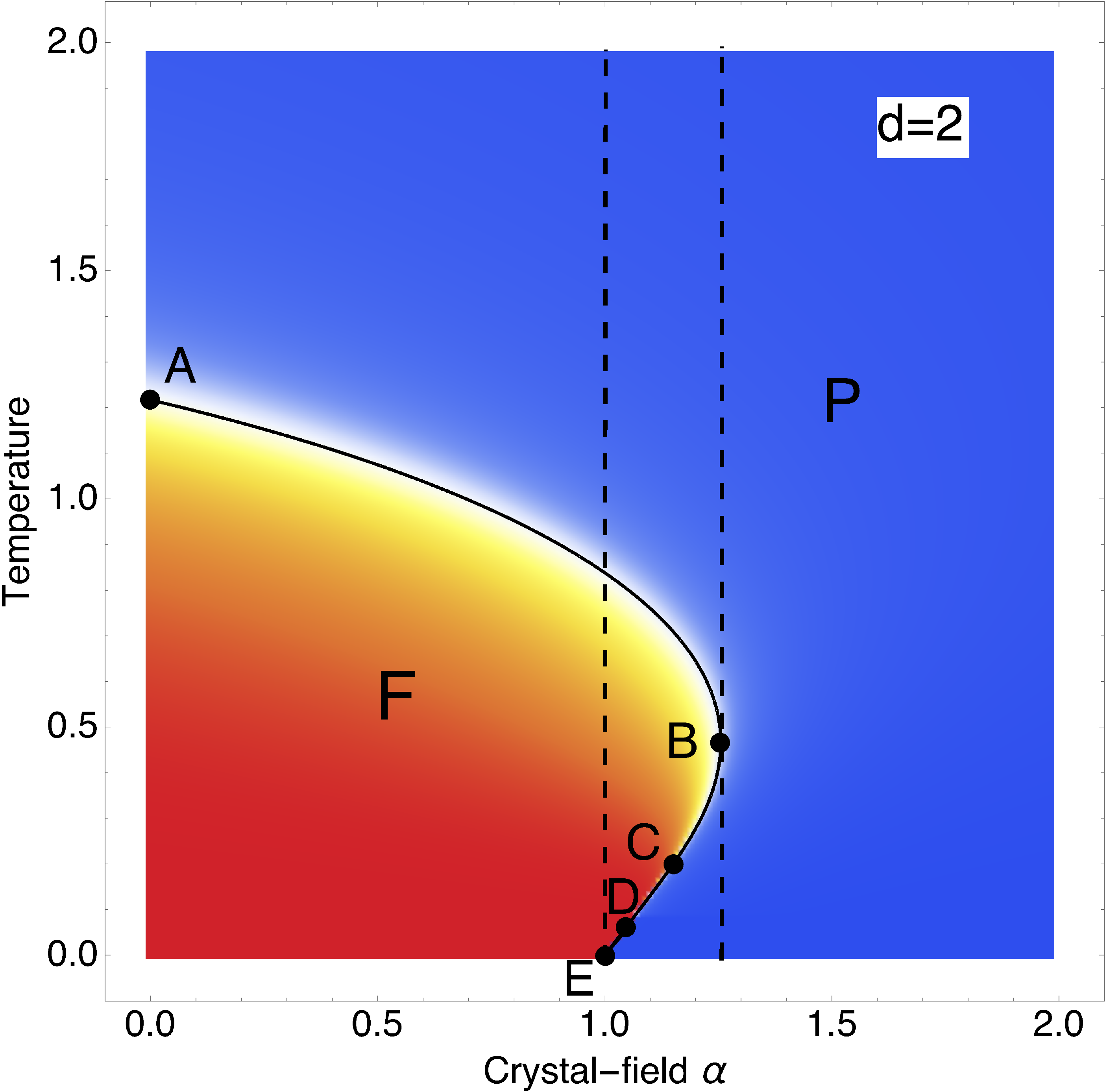}\
\includegraphics[width=8cm]{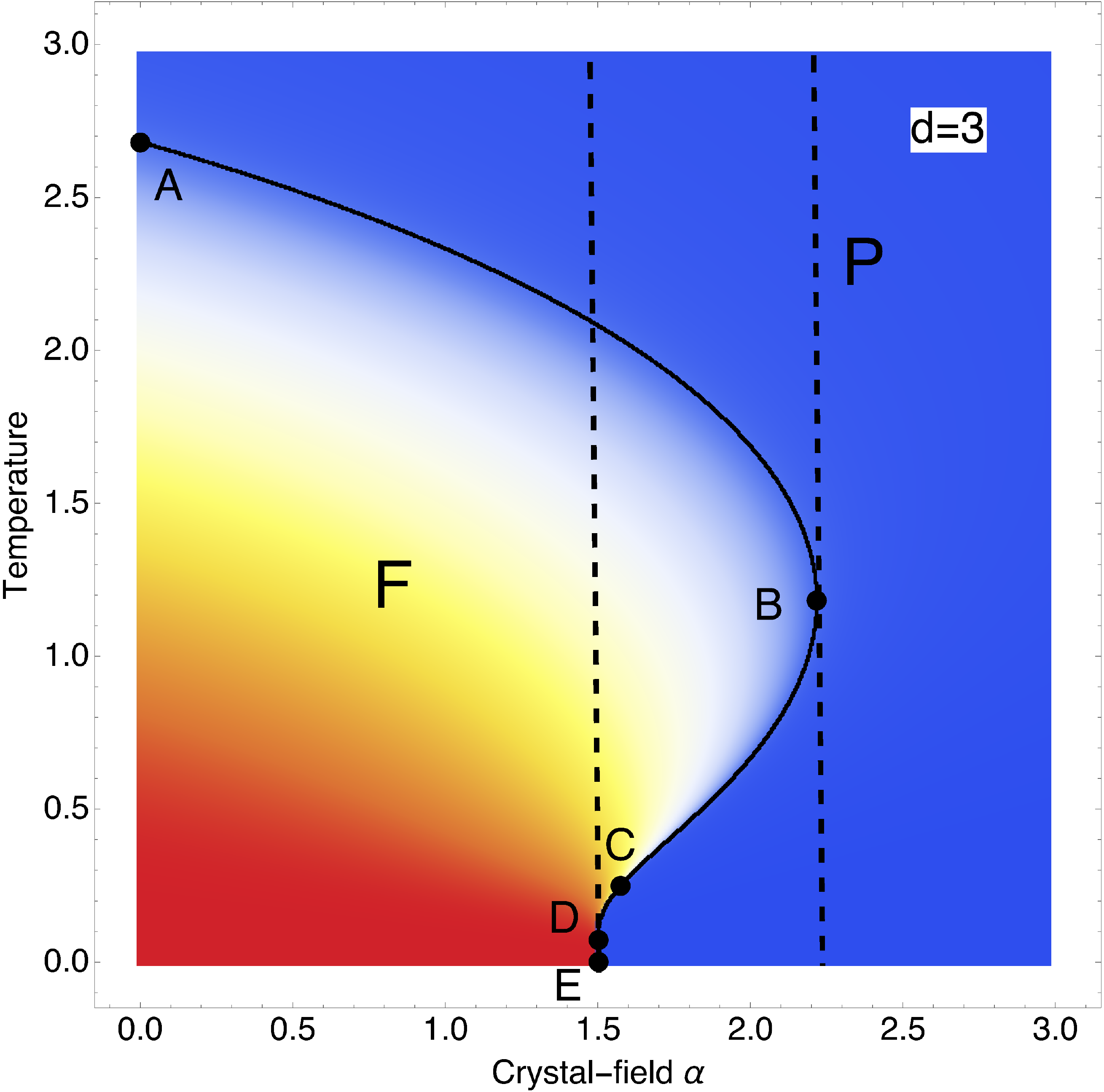}\\
\centerline{\includegraphics[width=8cm]{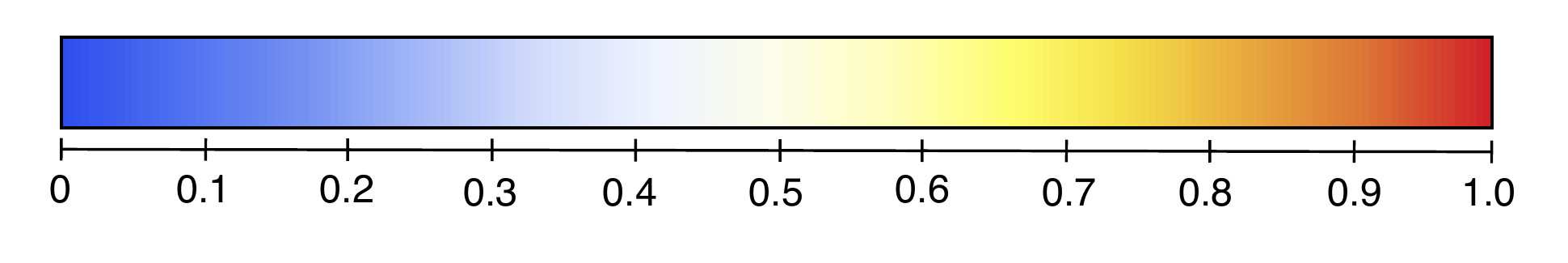}}
\caption{Temperature \emph{versus} crystal-field $\alpha$ phase diagram and magnetization density plot for the Blume-Capel model on DHL of dimension $d=2$ (left panel) and $d=3$ (right panel). F indicates the ferromagnetic region, and P labels the paramagnetic one for the respective dimensions. Letters A, B, C and D label the relevant points, which values are given in Table \ref{Tab1}. The dashed lines delimit the re-entrance regions as discussed in the text.
Colors represent the magnetization density following the ``temperature" color-palette on the bottom, from $0$ (\emph{cool} - blue) to $1$ (\emph{warm} - red).}
\label{Fig2}
\end{center}
\end{figure}

The phase diagram emerging from the renormalization flow in the $(J,\Delta)$ parameter space was drawn and studied in detail by the authors in \cite{rochaneto18}. An important feature in this phase diagram is the presence of a reentrant phase, also observed in the BC model defined in other nonregular lattices \cite{azhari22}, but not seen in numerical simulation studies of the BC model in hypercubic lattices (see, for instance, \cite{zierenberg17}). Here in  Figure \ref{Fig2}, we present the density plot of magnetization as a function of the temperature and crystal-field parameters, superimposed by the phase diagram that emerges from the real-space renormalization procedure. Panel (a) shows the plot for $d=2$ DHL while panel (b) shows the plot for $d=3$ DHL. In both plots, the ferromagnetic and paramagnetic regions (P) are indicated, while the relevant points $A,\, B,\, C,\, D$ and $E$ are presented in Table \ref{Tab1} and discussed below. 

\section{Quadrupole moment density order-parameter and its isothermal susceptibility}\label{sec3}
The thermodynamic behavior of the BC model can be exploited in any region of the phase diagrams shown in Figure \ref{Fig2}. In this work, however, we focus on the low-temperature region and more specifically the one close to the critical values of the crystal-field parameter $\alpha$ and around the relevant points $B$, $C$ and $D$ of the phase diagrams.  On this region and along the boundary that delimits the ferromagnetic and ordered-paramagnetic phase we shall find the locus of a tricritical point that marks the end of the continuous transition line followed by the first-order transition line, as predicted by the mean-field theories and observed through methods of numerical simulation applied to the BC model \cite{zierenberg17}. More specifically, we chose special points on the transition line where continuous and discontinuous transitions were observed in reference \cite{rochaneto18} through the behavior of magnetization density and the fraction of sites with spin state $S = 0$,  $n_0=(1-q)$, as defined below.
Table \ref{Tab1} provides the coordinates of these particular points for the model defined on lattices with $d=2$ and $d=3$.
\begin{table}[ht]
\renewcommand{\arraystretch}{1.5}
\setlength{\tabcolsep}{7pt}
  \centering 
\begin{tabular}{|c|ll|cc|cc|}
\hline
\multicolumn{1}{c}{}&\multicolumn{2}{c}{}  & \multicolumn{2}{c}{d=2 }& \multicolumn{2}{c}{ d=3 } \\ \hline
I& $T_\textrm I$& $-\infty $ &  $1.641$ & $ -\infty$& $3.830$ & $-\infty$ \\ \hline
A& $T_\textrm A$ & $ \alpha_\textrm A$  & $1.217 $ & $ 0.000$   & $ 2.682$ & $0.000$ \\ \hline
 B& $T_\textrm B$ & $ \alpha_\textrm B$ & $ 0.466$ & $1.255$ & $1.181$ & $2.218$ \\ \hline
C&  $T_\textrm C$ & $\alpha_\textrm C$&$0.200$ & $1.151$ & $0.260$ & $1.584$ \\ \hline
 D&  $T_\textrm D $ & $\alpha_\textrm D$&$0.060$ & $1.046$ & $0.070$ & $1.500$ \\ \hline
E &   $T=0$ & $\alpha^*$ &0 & 1.000& 0 & 1.500\\
\hline
\end{tabular}
  \caption{Coordinates of the relevant points in the phase diagram for the BC model in a DHL with dimensions 2 and 3: I=($T_\textrm I,\,-\infty$) spin 1/2 Ising model limit; A=($T_\textrm A,\; \alpha_\textrm A=0$) zero crystal-field  BC model; B=($T_\textrm B,\;  \alpha_\textrm B$) threshold point for absence of ferromagnetic phase  when $\alpha >  \alpha_\textrm B$; C=($T_{\textrm C},\; \alpha_{\textrm C}$) threshold point for a continuous phase transition for $T\geq T_{\textrm C}$ and D=($T_{\textrm D},\; \alpha_{\textrm D}$) threshold point for existence of phase transition with full discontinuity in the order parameters for $ T<T_{\textrm D}$ as explained in text below and D=($T=0,\; \alpha^*$) critical field for the zero temperature ferromagnetic-\emph{condensed} paramagnetic transition in the ground state.}
\label{Tab1}
\end{table}
\newpage
Our previous analysis of the behavior of the local magnetization and fraction $n_0$ of sites with spin-0 state performed in the reference \cite{rochaneto18} indicates that the locus of the tricritical point must be laid between the points C and D on the critical line shown in Figure \ref{Fig2}, in both cases $d=2$ and $d=3$, respectively.  In the following we further explore in detail the properties of $n_0$ at low temperatures in the phase diagram regions for $\alpha \geq \alpha^*$. 

\begin{figure}[ht]
\includegraphics[width=8cm]{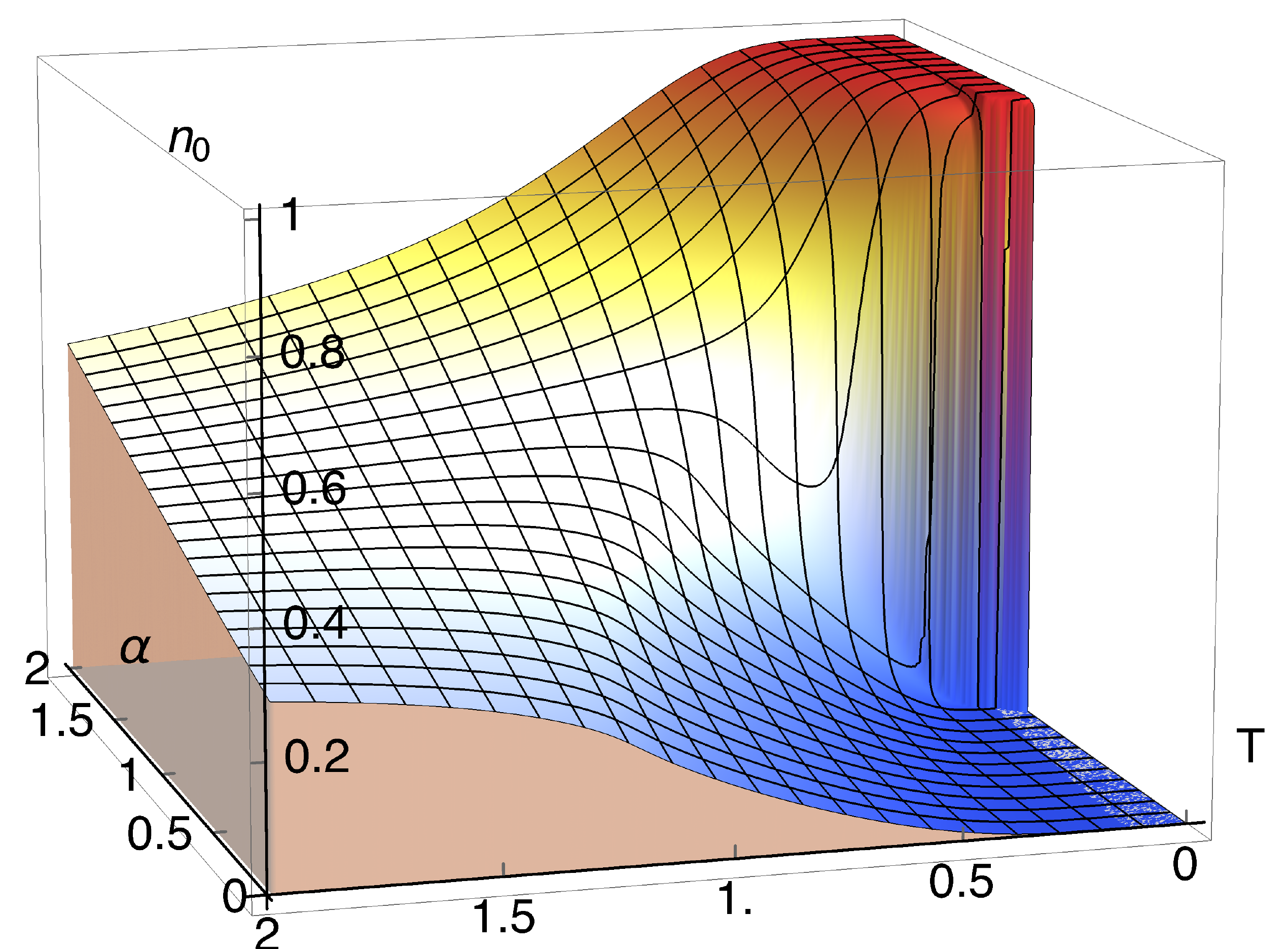}
\includegraphics[width=8cm]{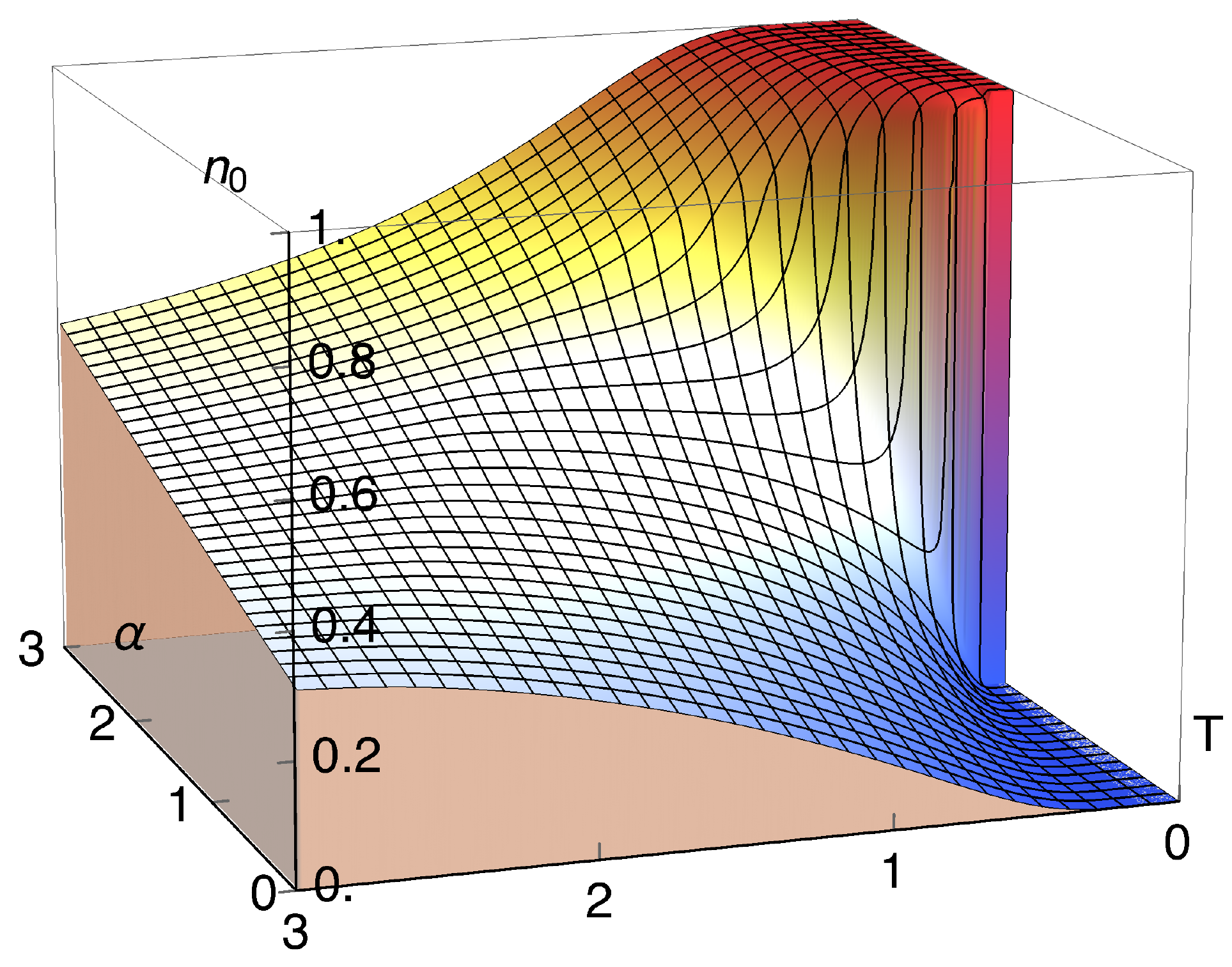}
\centerline{\includegraphics[width=8cm]{legenda.pdf}}
\caption{Fraction $n_0$ of sites with spin-state $S=0$ as a function of the Temperature $T$ and the crystal field $\alpha$ for the Blume-Capel model on DHL of dimension $d=2$ (left) and $d=3$ (right). Colors represent $n_0$ following the ``temperature" color-palette on the bottom from $0$ (\emph{cool} - blue) to $1$ (\emph{warm} - red).}
\label{Fig3}
\end{figure}
\subsection{Quadrupole moment density order parameter}\label{sec31}
 The fractions of sites $n_0$ and $n_\pm$, with spin states $S=0, \pm 1$, respectively, can be related as 
 \begin{equation}
 \begin{cases}
\ n_0+n_++n_-=1, \\
\ q=n_+ + n_-, \\
\ m=n_+ - n_- 
 \end{cases}
\quad  \therefore \qquad    
\begin{cases}
\ n_0 =(1-q), \\[1ex]
\  n_{\pm} =\displaystyle  \frac{1}{2}(q \pm m) 
\end{cases}   \label{eq7}
 \end{equation}
where the magnetization density $m$  and  the quadrupole moment density $q$ are defined in equation (\ref{eq3}). The ground state for zero magnetic field H, exhibits two possible configurations according to the value of the reduced crystal field $\alpha = D/K$. For $\alpha < \alpha^*$ the \emph{ferromagnetic phase} is stable and governed by the order parameter $m$ ($m=1$, $n_0=0$) associated with the field  $H$, while for $\alpha > \alpha^*$ the \emph{ordered-paramagnetic phase} has lower energy than the F-phase and is characterized by the order parameter $n_0$, ($m=0,\; n_0=1$), associated with the local crystal-field $D$. $\alpha^*$ is the critical value of the crystal-field strength that depends on the DHL dimensionality $d$. At $\alpha=\alpha^*$ ($T=0$) the spin system undergoes a first-order phase transition as we will show below. The limit regime  $\alpha \to -\infty$ corresponds to the suppression of the $S=0$ spin configurations and the model becomes equivalent to the $S=1/2$ Ising model.
 
The local values of $\langle S_i \rangle$ and $\langle S_i \rangle^2$ are temperature and crystal-field dependent and $m(T,\alpha)$ and $n_0(T,\alpha)$    can be exact and numerically calculated using the methodology developed in \cite{rochaneto18}. The behavior of the magnetization $m(T,\alpha)$, in the range of values where the phase transitions of interest occur, is illustrated in Figure 2 as a density   plot superimposed on the phase diagram obtained by the renormalization group transformations, for the cases of DHL's with $d=2$ and $d=3$ \cite{rochaneto18}.
  
Figure \ref{Fig3} displays the three-dimensional plots for the density $n_0(T,\alpha)$ as a function of the temperature $T$ and crystal-field parameter $\alpha$, for the $d=2$ and $d=3$ DHL's cases. The corresponding density plots of $n_0$  in the temperature \emph{versus} the crystal-field parameter space for $d=2$ and $d=3$ DHL's are depicted in Figures 6 and 7 of reference \cite{rochaneto18}, respectively. In Figure \ref{Fig3} the corresponding mesh lines, which can be used as a guide to analyzing the qualitative behavior of $n_0$ as a function of $T$ (or $\alpha$) were drawn by units of $0.1$ steps in both axes and plots. In both graphs, we observe an abrupt variation of $n_0(T,\alpha)$ at low temperatures in the range of values of $\alpha^*\leq \alpha \leq  \alpha_\textrm B$, which corresponds to the re-entrance regions delimited by the dashed lines drawn in the diagrams shown in Figure \ref{Fig2}. At low temperatures $n_0$ also characterizes the two distinct behaviors when $n_0\simeq 0$ corresponding to the F-phase while $n_0\simeq 1$ describing the OP-phase. On the contrary, at high temperatures $n_0 \to 1/3$ for every value range of $\alpha$ (corresponding to the paramagnetic phase in the diagrams of Figure \ref{Fig2}.

\begin{figure}[ht]
\includegraphics[width=8cm]{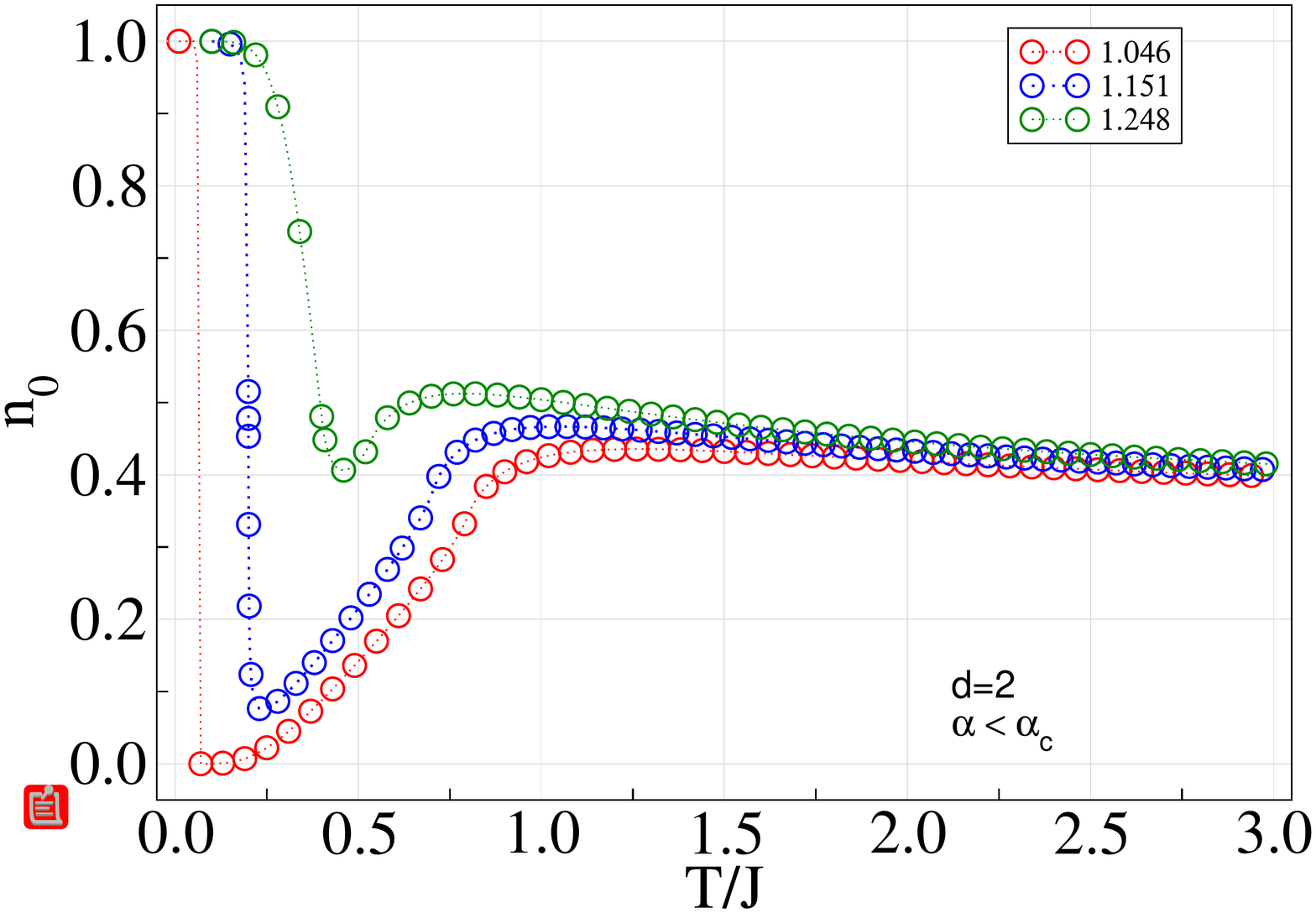}
\includegraphics[width=8cm]{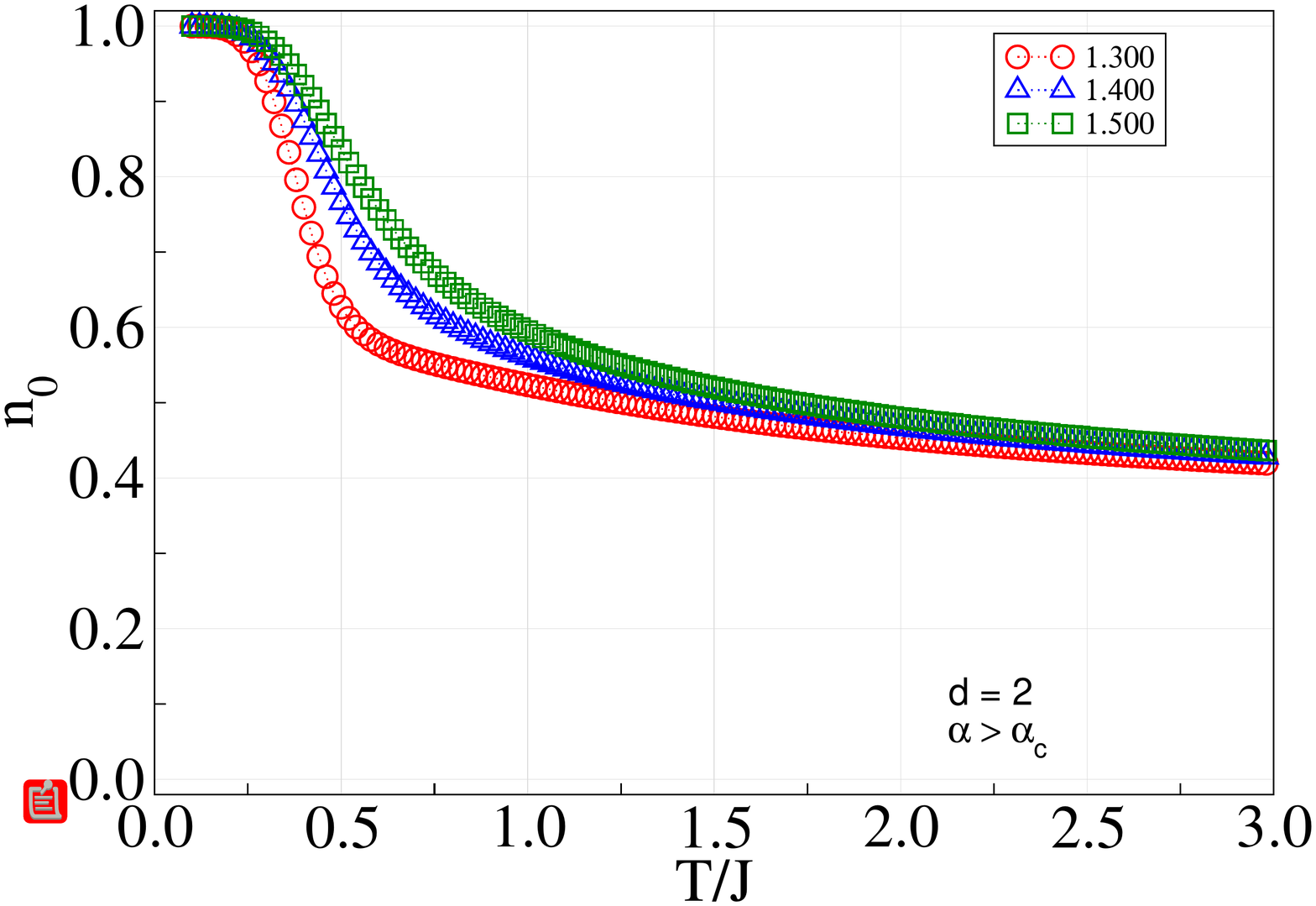}\\
\includegraphics[width=8cm]{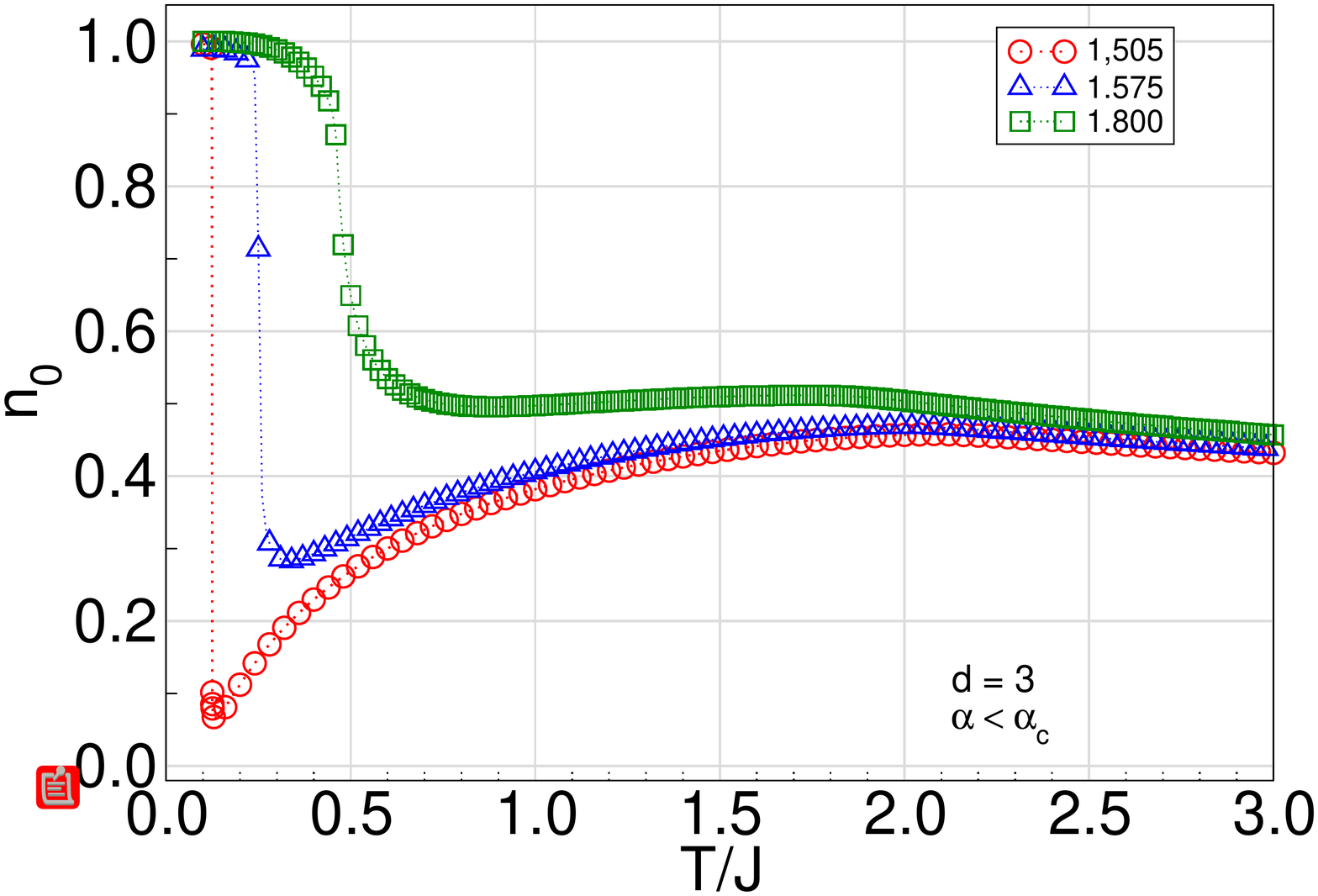}
\includegraphics[width=8cm]{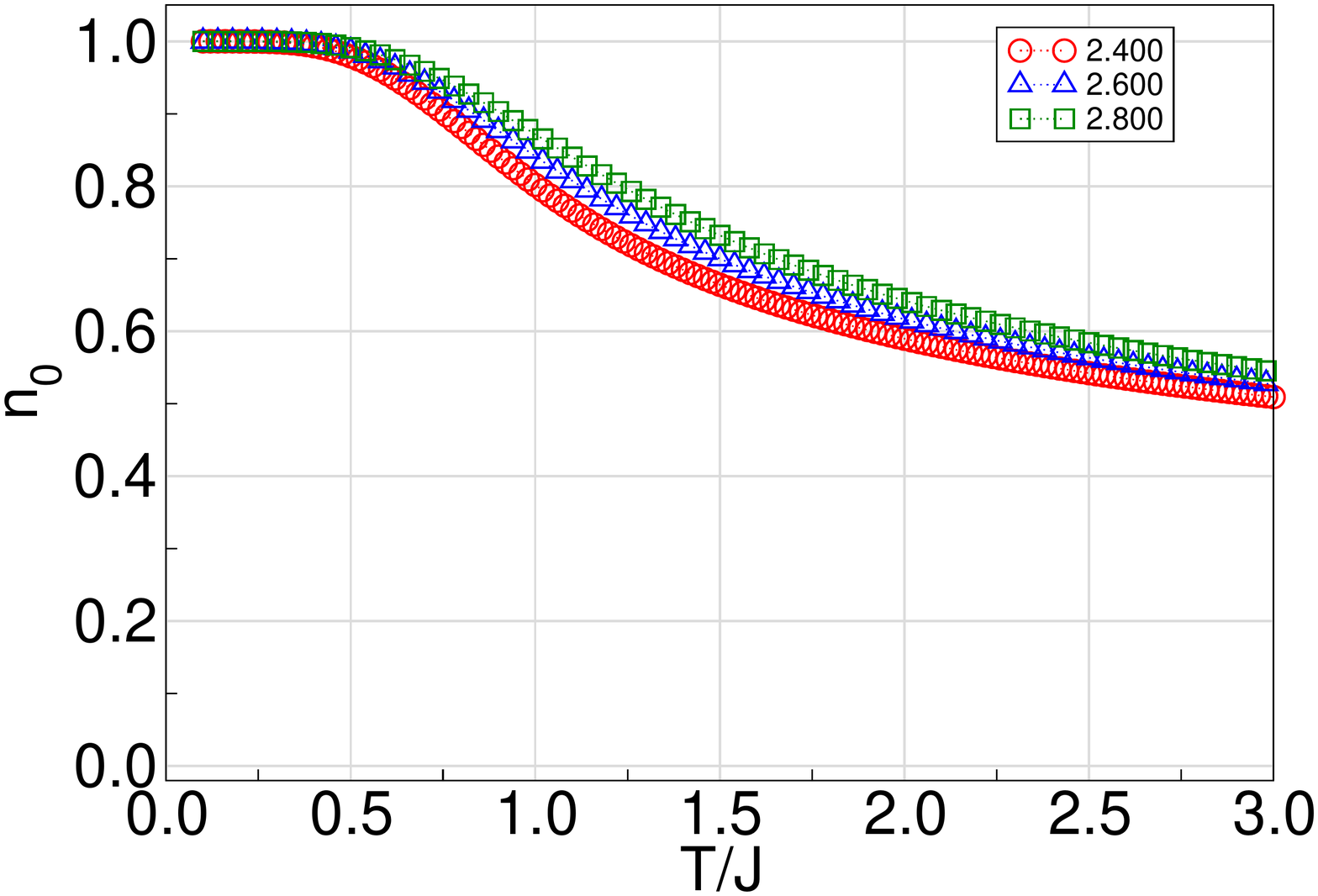}
\caption{Plots of the density $n_0$ as a function of the temperature for several values of the crystal-field parameter $\alpha$ for the BC model defined on a $d=2$ (upper panels) and $d=3$ (lower panels) DHL. In both cases, the left panels explore the plots for values of  $\alpha* < \alpha < \alpha_\textrm B$ (reentrant region), while the right panels depict the plots for  $\alpha >\alpha_\textrm B$, respectively, according to the legends.}
\label{Fig4}
\end{figure}
To explore, in more detail, the behavior of $n_0$ at low temperatures, we exhibit in Figure \ref{Fig4} some particular plots of $n_0$ as a function of the temperature for fixed values of the crystal-field parameter $\alpha$ for the model defined on DHL's with $d=2$ (upper panels) and with $d=3$ (lower panels). In both cases the left pannels shows the plots for $\alpha_\textrm D \leq \alpha \leq \alpha_\textrm B$ (within reentrance interval) for three values of $\alpha$, while the right ones display the plots for values of $\alpha > \alpha_\textrm B$, that is, away the reentrance region, for three values of $\alpha$ according to the legend. Such plots correspond to particular cases of the mesh lines along the temperature axis in each case  shown in Figure \ref{Fig3}. In both cases  ($d=2$ and $d=3$), the abrupt drop followed by a pronounced minimum, in the reentrant region and at low temperatures, reflects the ingress of the system on the ferromagnetic phase region as the temperature is raised, as it is also evident in the phase-diagrams displayed in Figure \ref{Fig2}. In the upper-left panel the particular plots for $\alpha=\alpha_\textrm D=1.046$ and $\alpha=\alpha_\textrm C=1.151$ there exist dramatic drop of $n_0$ at very low temperatures signalizing a \emph{first-order} transition from the ordered-paramagnetic to the ferromagnetic phase, while for $\alpha=1.248<\alpha_\textrm B$  a \emph{continuous} transition occurs from the ferromagnetic phase to the paramagnetic one.

Similar behavior also occurs for the case $d=3$, as shown in the lower left panel for values of $\alpha=1.505$ and $1.575$ (first order) and $1.800$ (continuous).
In contrast, the right panels of Figure \ref{Fig4} illustrate the plots for values of $\alpha$ beyond $ \alpha_\textrm B$ where the behavior of $n_0$ becomes monotonically decreasing from $n_0=1$ to $n_0=1/3$, as the temperature is augmented from low to high values, for a fixed the values of $\alpha$, for both cases of $d=2$ (upper) and $d=3$ (lower) DHL's.   However, such decreasing course exhibits a concavity signal changing signalizing a continuous \emph{soft-passage} from the ordered-paramagnetic configuration to the paramagnetic phase, as we discuss in the following subsection.

\begin{figure}
\begin{center}
\includegraphics[width=8.1cm]{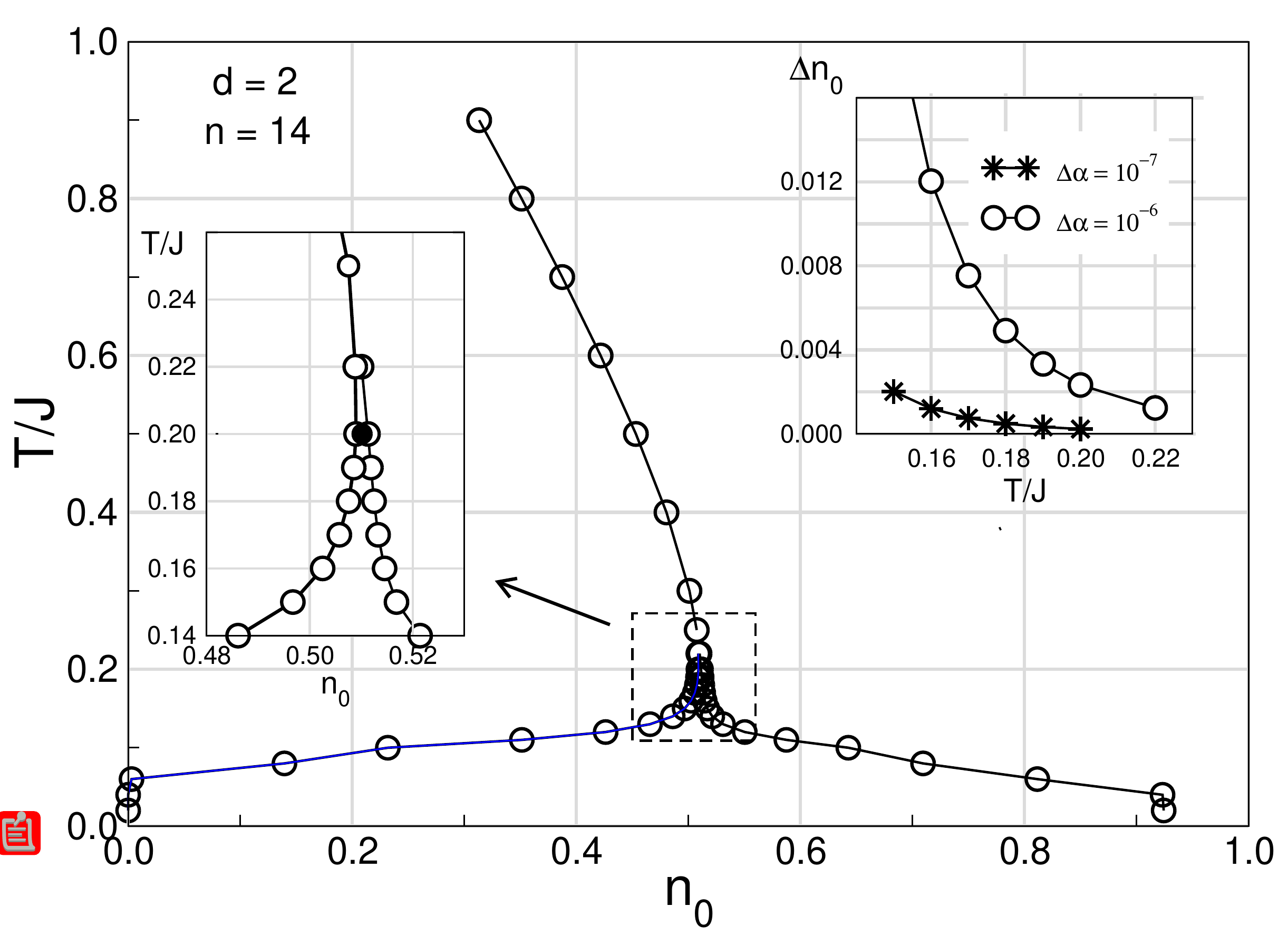}
\includegraphics[width=8.1cm]{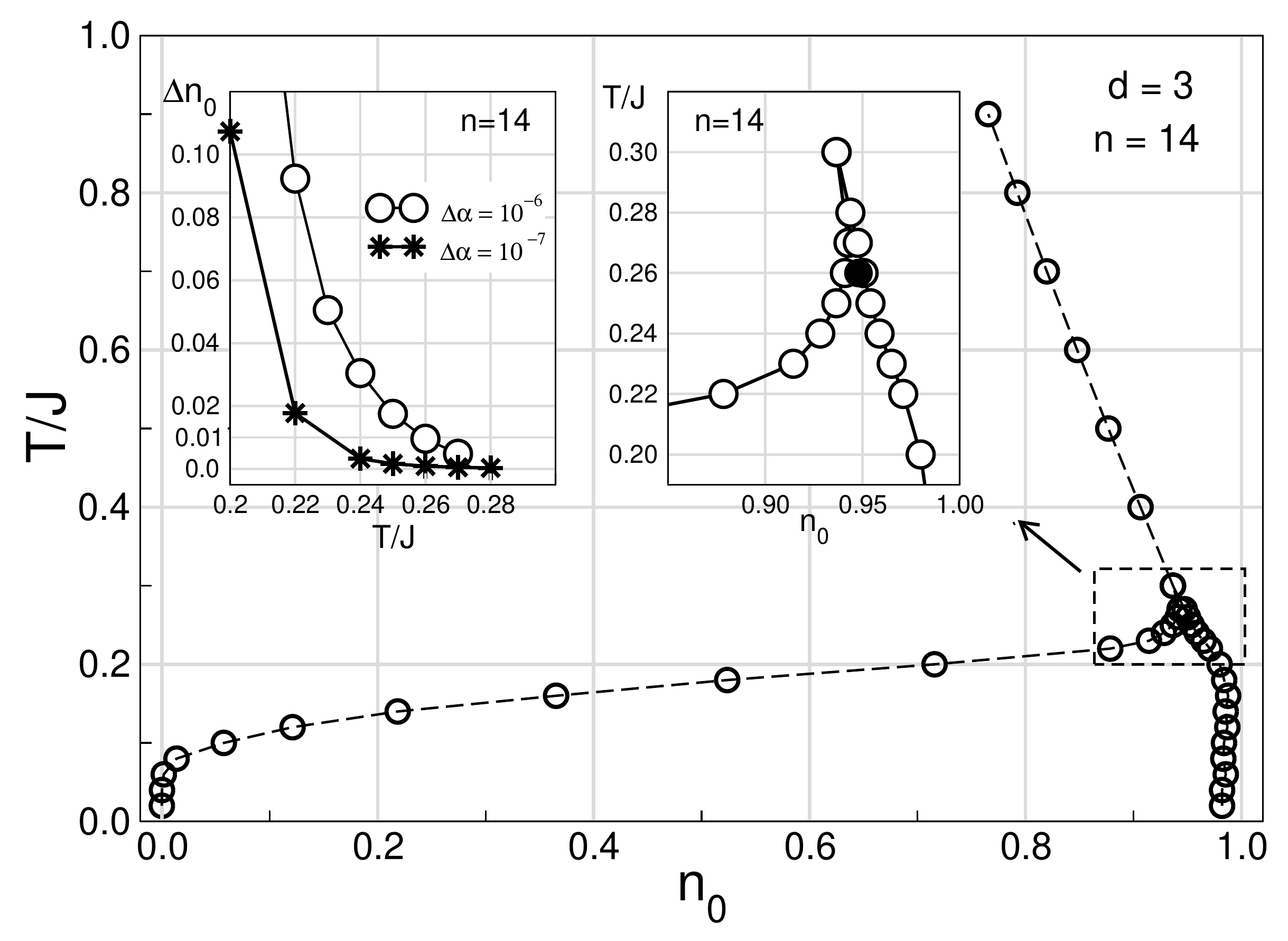}
\caption{Phase diagram temperature as a function of fraction $n_0$ for DHL's with $n=14$ generations: $d=2$ (left) and $d=3$ (right). The insets indicated by the arrows show the magnification of the region delimited by the dashed rectangles, where a solid black circle marks the possible locus of the corresponding tricritical points (TCP) with $T_\textsf{TCP}=0.20(1),\, \alpha_\textsf{crit}=1.150882(1)$,  $n_0=0.510163845$ for $d=2$,  and $T_\textsf{TCP}=0.26(1)$,  $\alpha_\textsf{crit}=1.583952(1)$, $n_0=0.947887$ for $d=3$. Other insets display the variation of $n_0$ as a function of the temperature below the corresponding TCP and for two values of $\Delta \alpha= \alpha_+-\alpha_-=10^{-6}$ and $10^{-7}$.}
\label{Fig5}
\end{center}
\end{figure}

To characterize the first-order transition evidenced by the rude behavior of $n_0$ at very low temperatures, in the reentrant regions of the phase-diagrams, we analyze the plots of $n_0$ as a function of the temperature \emph{along} the \emph{ferromagnetic-paramagnetic} transition line. In Figure \ref{Fig5} we plot $n_0(\alpha_\pm)$, for a fixed temperature $T=T_\textsf{crit}$, for  values of the $\alpha_\pm= \alpha_\textsf{crit} \pm 1 \times 10^{-6}$, i.e. immediately before  and after the transition line whenever they differ, for the model on $d = 2$ and $d = 3$ DHL. These diagrams reproduce the experimental phase diagram for mixtures of liquid ~${^3}$He --${^4}$He, along the $\lambda$-line and close to the two-fluid critical mixing point, which is well described by the spin-one Blume-Emmery-Griffiths model \cite{griffiths70,Blume1971}, where the corresponding density $n_0$ mimics the fraction of atoms ${^3}$He.

In both plots, for $d=2$ (left panel) and $d=3$ (right panel) DHL's, the presence of the tricritical point is marked by a black solid symbol in the corresponding insets indicating that at low temperatures below this point, the system undergoes a first-order transition with the coexistence of a nonzero (within the numerical precision) fraction of $n_0$ and the remaining fraction with spins at states $S=\pm 1$ ($q\neq 0$). The two insets in both panels explore the behavior of $n_0$ in the vicinity of the TCPs' locus. One of the inserts in each Figure \ref{Fig5} shows an enlargement of the TCP location region, as indicated by the arrow, marking with a solid black symbol the expected position of the TCP. The second inset shows the behavior of $\Delta n_0=n_0(\alpha_+)-n_0(\alpha_-)$ as a function of temperature for two possible situations of the degree of precision of $\Delta \alpha=\alpha_+ - \alpha_-$, as indicated in the legend. This analysis indicates that the TCP is approximately located at the values corresponding to point C of the phase diagrams displayed in Figure \ref{Fig2}, which are at $T_\textsf{crit} = 0.20(1)$  and $T_\textsf{crit} = 0.26(1)$, for $d = 2$ and $d = 3$ DHLs, respectively. We consider both $\Delta \alpha\sim 10^{-6}$ and $10^{-7}$ to obtain the values $T_\textsf{TCP}$ with uncertainty of order $10^{-2}$ setting the precision limits for $\Delta n_0$ in $10^{-3}$ and $10^{-2}$ for the cases of DHL with $n=14$ generations and dimensions $d=2$ and $d=3$, respectively. It is important to emphasize here that the precision in determining the numerical values of the tricritical point can be improved if the numerical calculation algorithms are implemented with the aid of more powerful computational resources. The position of the tricritical point lies in  the reentrant region on the frontier of the ordered phase as here obtained and below confirmed by susceptibility/specific heat rescaling studies, a result that differs from the one observed in the BC model defined in other non-regular lattices \cite{azhari22} where the locus of TCP was found at the extreme of the reentrant phase. We also find it relevant to mention here that, as the temperatures is lowered below TCP, the multifractal spectrum of the local magnetization evolves narrowing until it reaches a purely fractal behavior (single point), as discussed in the reference \cite{rochaneto18}. 
In the following subsections, we explore the thermodynamical properties of $n_0$.
\subsection{Isothermal quadrupole moment susceptibility}\label{subsec31}
The isothermal quadrupole moment susceptibility associated with the density $n_0$ with respect the crystal-field $\alpha=D/K$, which can be viewed as the conjugate field associated to the order-parameter $n_0$, can be defined by 
\begin{equation}
\label{eq8}
\chi_T= \frac{\partial n_0}{\partial D}\Big|_{T}=\frac{1}{K} \frac{\partial n_0}{\partial \alpha}\Big|_{T}= - \dfrac{1}{K}\frac{\partial q}{\partial \alpha}\Big|_{T}\ .
\end{equation}

\begin{figure}
\includegraphics[width=80mm]{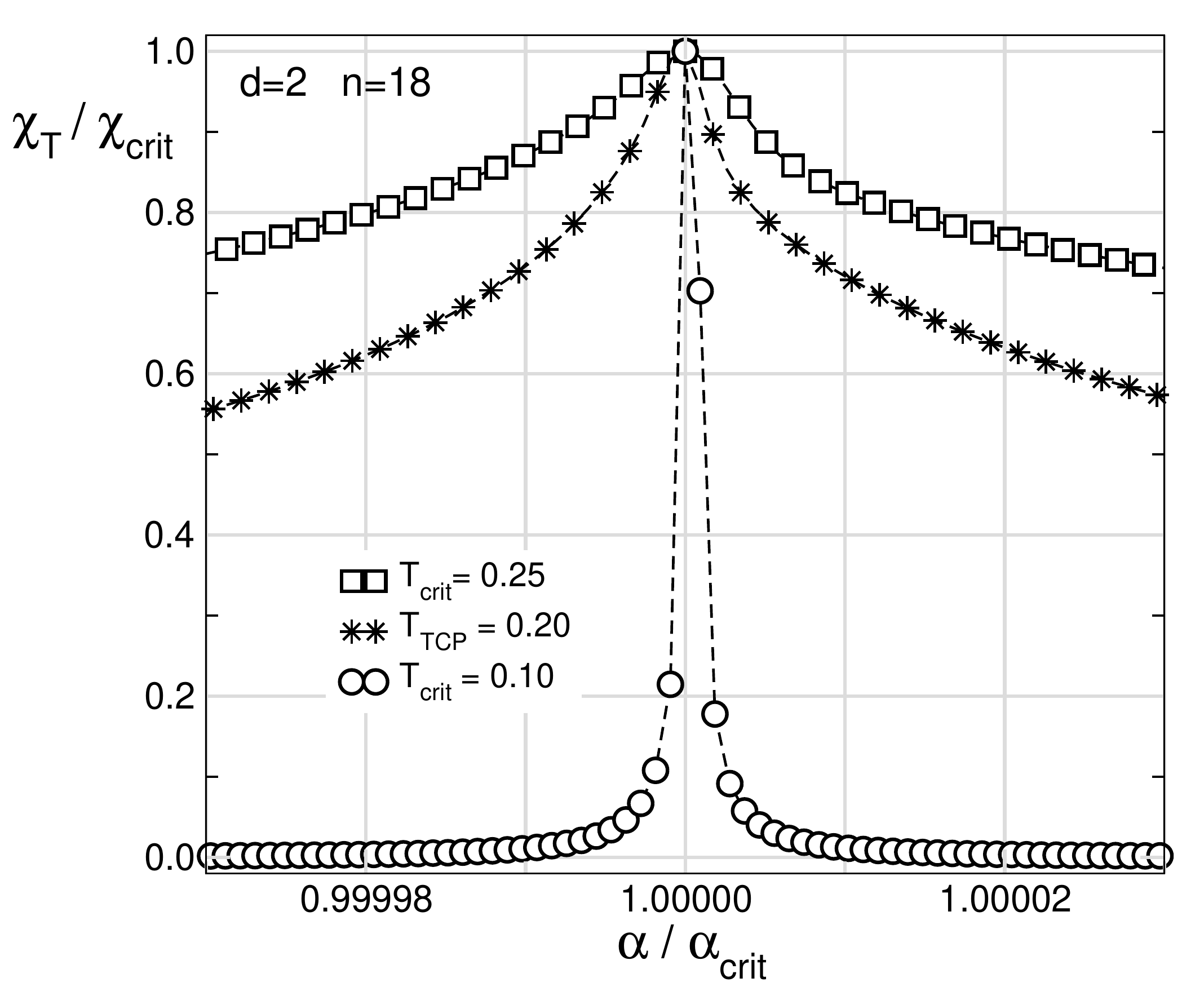}
\includegraphics[width=84mm]{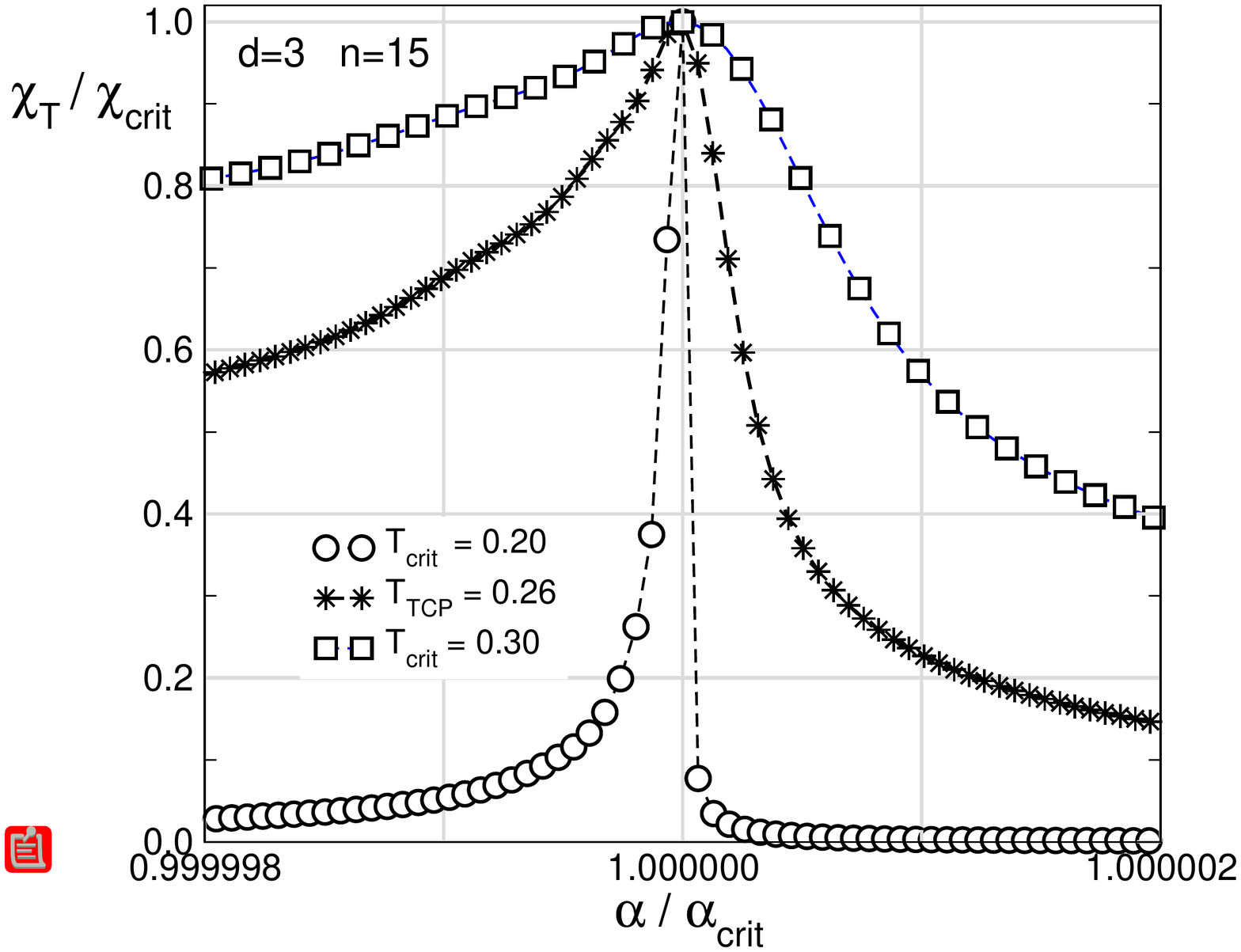}
\caption{Normalized quadrupolar susceptibility $\chi_T/\chi_\textsf{crit}$ as a function of the reduced crystal field $\alpha/\alpha_\textsf{crit}$ for fixed temperatures below, at and above the tricritical point. Left panel: for $d=2$ DHL's with $n = 18$ generations drawn for critical temperatures $T_\textsf{crit}=0.15,\; 0.20\, \textsf{(TCP)}$ and $0.26$. Right panel: for $d=3$ DHL's with $n = 15$ generations drawn for critical temperatures $T_\textsf{crit}=0.20,\; 0.26\, \textsf{(TCP)}$ and $0.30$ according to the legends.} 
\label{Fig6}
\end{figure}

Figure \ref{Fig6} illustrates the behavior of $\chi_T$ as a function of the crystalline field $\alpha$, for fixed temperatures below, at and above the corresponding TCP. The graphs were obtained by numerical derivative of $n_0$ for a DHLs with $d=2$ (left panel) and $d=3$ (right panel) and drawn in normalized scales for comparison purposes. We observed in both cases the presence of a very sharp peak in the associated critical value of $\alpha_\textsf{crit}(T)$ for temperatures below the tricritical point, corroborating the expected behavior for first-order transitions in finite systems. In general, for first-order transitions in infinite-system limit a divergence in the susceptibility is not expected but an asymmetric finite peak due to the order-parameter gap at the critical point. The $\chi_\textsf{T}$ plot for $T=0.10$ (left panel) shows a reminiscent behavior of this peak for a DHL  with $d=2$ ($n=18$),  while for $T=0.20$ (right panel) this asymmetric pattern is unambiguous for a $d=3$ ($n=15$) DHL. Furthermore, the peaks at and above the tricritical temperature become finite and much wider as observed in continuous transition for finite systems. In addition, we found in all cases that the peaks present a certain asymmetry, predominantly more accentuated in the case of the $d=3$ DHL.

In this work we deal with exact thermodynamic quantities which can be numerically calculated in finite DHLs. Therefore the characterization of the nature of the phase transition in the vicinity of the tricritical point can be   explored quantitatively through a finite-size scaling analysis (FSS) \cite{binder92,newman99} of the dependence of the maximum value of $\chi_T$, namely $\chi_T^{[\max]}$, as a function of the size $L=2^n$ of the $n$-generation DHL. 

\begin{figure}
\begin{center}
\includegraphics[width=8.2cm]{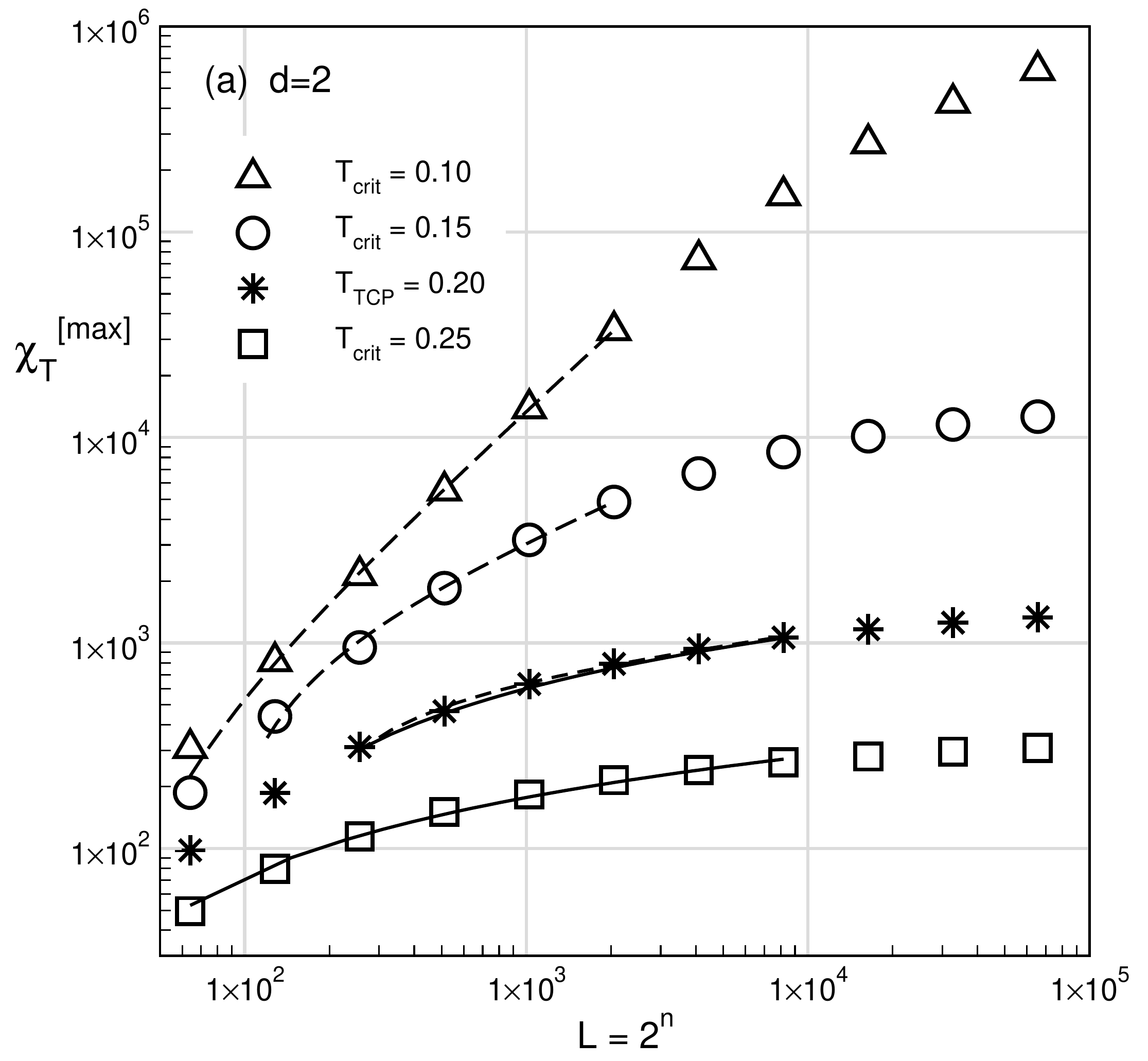}
\includegraphics[width=8.1cm]{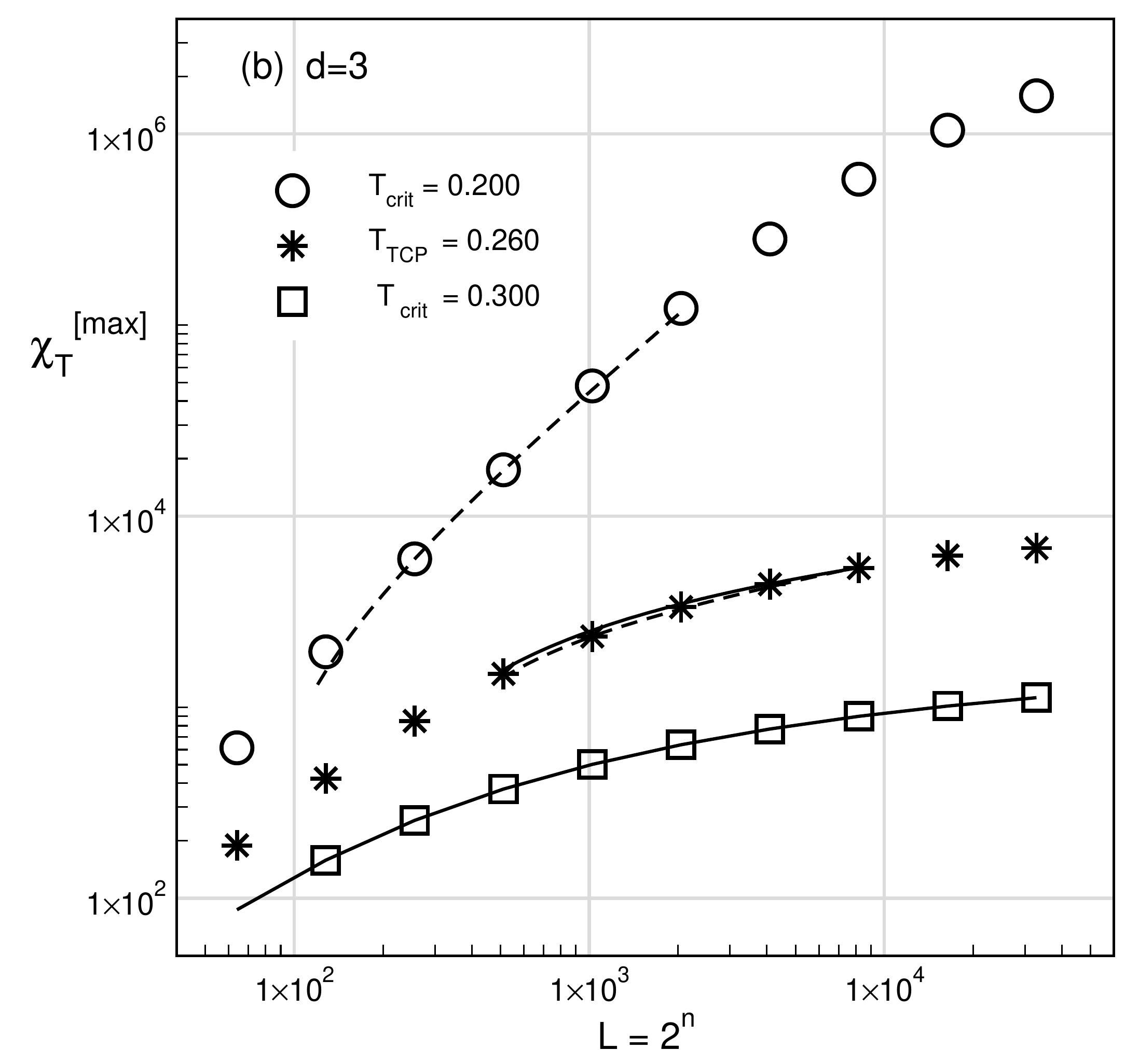}
\caption{Log-Log plots of the values of the quadrupolar susceptibility peaks $\chi_\textsf{T}^{[\max]}$ as a function of the DHL-size $L$ calculated on the F-P transition boundary and exhibiting the corresponding fitting interval according to the legends and parameters given in Table \ref{Tab2}. Panel (a) for $d=2$-DHL, at  $T_\textsf{crit} = 0.10$ (triangle), $0.15$ (circle), $T_\textsf{TCP}=0.20$ (star) and $T_\textsf{crit} =0.25$ (square). Panel (b), for $d=3$-DHL, at $T_\textsf{crit}= 0.200$ (circle), $T_\textsf{TCP}=0.26$ (star) and $T_\textsf{crit} =0.30$ (square).  In both panels, the dashed lines indicate the adjustment with a power law  -- with first-order correction -- and the solid lines indicate the logarithmic fitting, all plotted within the interval where the fit was performed.}
\label{Fig7}
\end{center}
\end{figure}

In general, it is expected that the peak value $\chi_T^{[\max]}$ increases, and its position $\alpha_{\textsf{crit}}(L)$ shifts as the size of the lattice increase. Following Zierenberg \emph{et al} \cite{zierenberg15} we proceed for the FSS-analysis of $\chi^{\max}_T(L)=\chi_T(\alpha_{\textsf{crit}}(L))$ considering temperatures slight above, slight below and at the tricritical value, namely $T_\textsf{TCP}$, for DHLs with dimension $d=2$ and $d=3$. We made a detailed study of the fitting curves on the lattice size $L=2^n$ intervals from $n=6$ to $n=16$,  for each fixed temperature. We   searched the intervals where the fitting curve to the data reached reasonable values of accuracy considering the R-squared of the regression as the control parameter, which is appropriated to compare rival models from a independent-variable specific sample \cite{hagquist98}. For  sake of comparison we chose the threshold of $R^2\geq 99.9\%$ as the criterion for a good accuracy and identify two distinct regimes for $\chi_t^{[\max]}(L)$ according the phase-transition is of first-order or continuous, as displayed in the plots shown in Figure \ref{Fig7}. Panel (a) illustrates these behaviors for $T_{\textsf{crit}}=0.10$ (triangles), $0.15$ (circles), $T_{\textsf{TCP}}=0.20$ (stars) and $T_{\textsf{crit}}=0.25$ (squares) for a $d=2$-DHL. Panel (b) do it for $T_{\textsf{crit}}=0.20$ (circles), $T_{\textsf{TCP}}=0.26$ (stars) and $T_{\textsf {crit}}=0.30$ (squares) for a $d=3$-DHL. Plots of data are drawn in log-log scales to accommodate the whole range of values and also display the best fitting curves. 
Two distinct behaviors emerge from this analysis: 
\begin{description}
    \item[(a)] For first-order transitions ($T_\textsf{crit} < T_\textsf{TCP}$), the best fits are described by power-law functions with first-order corrections. i.e.
    $$\chi^{[\max]}\sim a L^p (1+ a'L^{-1}).$$
     \item[(b)] For continuos transitions ($T_\textsf{crit} > T_\textsf{TCP}$), the best fits follow a logarithmic behavior,
     $$
     \chi^{[\max]}\sim b+ b'\textrm{Log} (L).
     $$
     \item[(c)]  For the transition at $T_\textsf{crit} = T_\textsf{TCP}$ both the power-law function and the logarithmic behavior is well accepted considering within the same adopted size scaling range and the numerical criterion of best fit.
     \end{description}
Table \ref{Tab2} displays for each temperature the adjustment coefficients $a$, $a'$, $p$, $b$ and $b'$, their corresponding scaling intervals and the $R^2$ parameter, respectively. It is important to note here that specifically at the estimated values for $T_\textsf{TCP}$ both type of fitting curves were possible with good fitting coefficients $R^2$ of the same order of magnitude but within a smaller size intervals $L$, as shown in Figure \ref{Fig7} and in Table \ref{Tab2}. 
\begin{table}[ht]
\renewcommand{\arraystretch}{1.2}
\setlength{\tabcolsep}{5pt}
  \centering 
\begin{tabular}{|c|c|c c c  |c c|c|c|}
\hline
$d$ & $T_\textrm{crit}$ &$a$ & $p$  & $a'$ & $b$ & $b'$  & R-squared. & $L=2^n$ interval \\ \hline  
  & $0.100$ &$2.4(3)$ & $1.25(1)$ & $-30(10)$ & $\cdots$ &$\cdots$ & $0.999978$ & $n=6 \to 11$ \\   
  & $0.150$ &$52(15)$ & $0.60(4)$ & $-75(15)$ & $\cdots$ &$\cdots$ & $0.999577$ & $n=7 \to 11$ \\   
  $2$ & $0.200^\star$ & $180(30)$ &$0.20(2)$ &$-114(14)$ & $\cdots$ & $\cdots$   & $0.999692 $ & $n=8 \to 13$ \\ 
  & $0.200^\star$ & $\cdots$ &$\cdots$ &$\cdots$ & $-900(40)$ & $500(11)$ & $0.999758$ & $n=8 \to 13$ \\  
  & $0.250$ & $\cdots$ &$\cdots$ &$\cdots$ & $-135(8)$ & $104(3)$ & $0.999924$ & $n=8 \to 13$ \\ 
    \hline  
  & $0.200$ & $6(1)$ & $1.30(2)$ & $-68(17)$ & $\cdots$ & $\cdots$   & $0.999692 $ & $n=7 \to 11$ \\   
$3$ & $0.260^\star$ & $340(60)$ & $1.31(2)$ & $-200(20)$ & $\cdots$ & $\cdots$   & $0.999882 $ & $n=9 \to 13$ \\ 
    & $0.260^\star$ & $\cdots$ & $\cdots$ & $\cdots$ & $-7700(300)$ & $3400(100)$   & $0.999614 $ & $n=9 \to 13$ \\  
 & $0.300$ & $\cdots$ & $\cdots$ & $\cdots$ & $-730(20)$ & $410(5)$.   & $0.999759 $ & $n=7 \to 15$ \\ \hline 
\end{tabular}
  \caption{Adjustment coefficients  $a$, $a'$, and $p$ for the power-law fitting $\chi^{[\max]}(L) = a L^p (1+ a'L^{-1})$, $b$ and $b'$ for the logarithmic fitting  $\chi^{[\max]}(L) = b + b'\ln L $ and the associated $R^2$-parameter for the plots shown in Figures \ref{Fig7}(a) and 7(b) for DHL's with $d=2$ and $d=3$, respectively. Last column indicates the $n$-generation interval for the size scaling region $L=2^n$.  Symbol $\star$ marks the above estimated values of $T_\textsf{TCP}$. Standard deviations of the coefficients are indicated between parentheses.}
\label{Tab2}
\end{table}
\section{Internal Energy and Specific Heat}\label{sec4}
In this subsection, we pay special attention  to the internal energy and constant crystal-field specific heat defined from the heat capacity defined by $C_D=(\partial \mathcal E/\partial T)_D$, $\mathcal E=\langle H_\textsf{BC} \rangle -DQ$ being the enthalpy.  In the present model $C_D$ is associated with the \emph{specific heat} at constant crystal-field and zero magnetic  field, $c_\alpha$, given by
\begin{equation}
\label{eq9}
c_\alpha= \frac{\partial u}{\partial T}\Big|_{\alpha},
\end{equation}
where the average exchange energy per bond $u$ for a $n$-generation DHL defined by equation (\ref{eq4}) is given by
\begin{equation}
\label{eq10}
u=\frac{1}{N_b^{(n)}}\sum_{\langle ij \rangle}\langle S_i S_j \rangle.
\end{equation}
$N_b^{(n)}$ is the total number of bonds in a DHL with $n$ generations.
 The  temperature and crystal-field dependent exact local quantities $ \langle S_i S_j \rangle$ associated with the $\langle ij \rangle$ bond are numerically obtained by  the recursion procedure derived in reference \cite{rochaneto18}.

\begin{figure}
\begin{center}
\includegraphics[width=8cm]{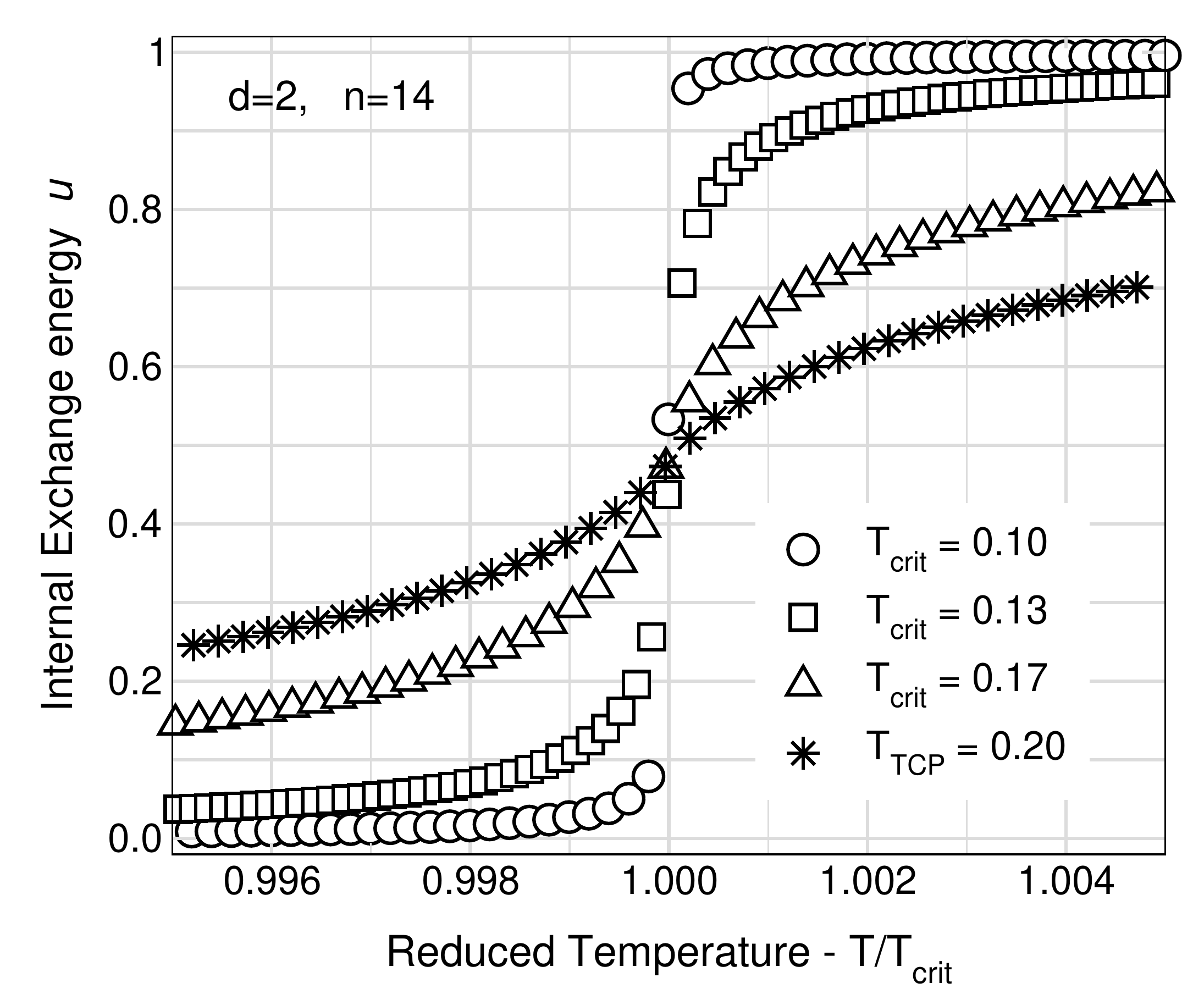}
\caption{Internal reduced energy density (in units of $K$) as a function of $T/T_\textsf{crit}$ for four fixed values of $\alpha_\textrm D< \alpha< \alpha_\textrm C$ for a $d = 2$ DHL with $n = 14$ generations. The symbols label the values of the respective critical temperatures according to the legends.}
\label{Fig8}
\end{center}
\end{figure}

Figure \ref{Fig8} illustrates the low-temperature dependence of $u$ when crossing the phase diagram of Figure \ref{Fig2} along a vertical line for fixed values of  $\alpha$ in the interval $\alpha_\textrm D \leq \alpha \leq \alpha_\textrm C$, calculated for a $d=2$ DHL with $n=14$ generations. In this re-entrant region and at low and decreasing temperatures occurs the F-OP transition from the ferromagnetic phase to the ordered-paramagnetic phase, the latter dominated by spins configurations with $S = 0$ state. In this transition the internal energy density $u$ undergoes sharp variations below the critical temperature for values of $1 < \alpha_\textsf{crit} < 1.15$, as illustrated for four fixed values of the $\alpha_\textsf{crit}$ parameter in this range, corresponding respectively to the critical temperatures $0.10,\; 0.13,\; 0.17$ and $T_\textsf{TCP} =0.20$, as indicated in the figure legends. 

As the critical crystal-field parameter $\alpha_\textsf{crit}$ is lowered along the ordered F-OP transition frontier (from the points C $\to$ D in the left panel of Figure \ref{Fig2}), the behavior of the energy changes drastically from smooth to abrupt signalizing a presence of the tricritical point connecting the second-order transition line with the first-order one along the such frontier, as above seen. This result also corroborates the indications suggested by the authors in \cite{rochaneto18} when studying the multifractal behavior of the local magnetization of the BC model. Similar results were also obtained for the case of DHL with $d = 3$.

\begin{figure}[ht]
\begin{center}
\includegraphics[width=8cm]{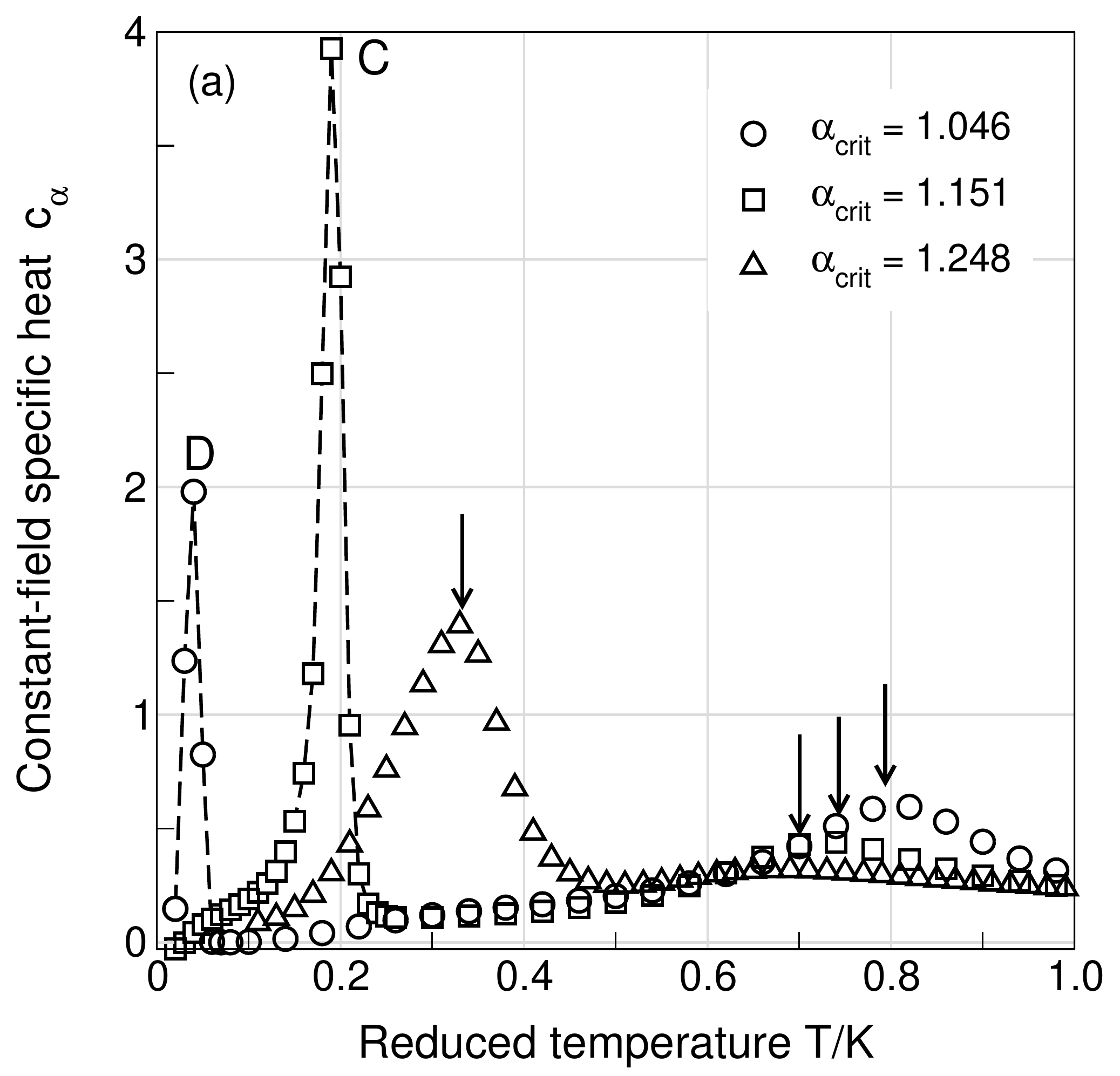}
\includegraphics[width=8cm]{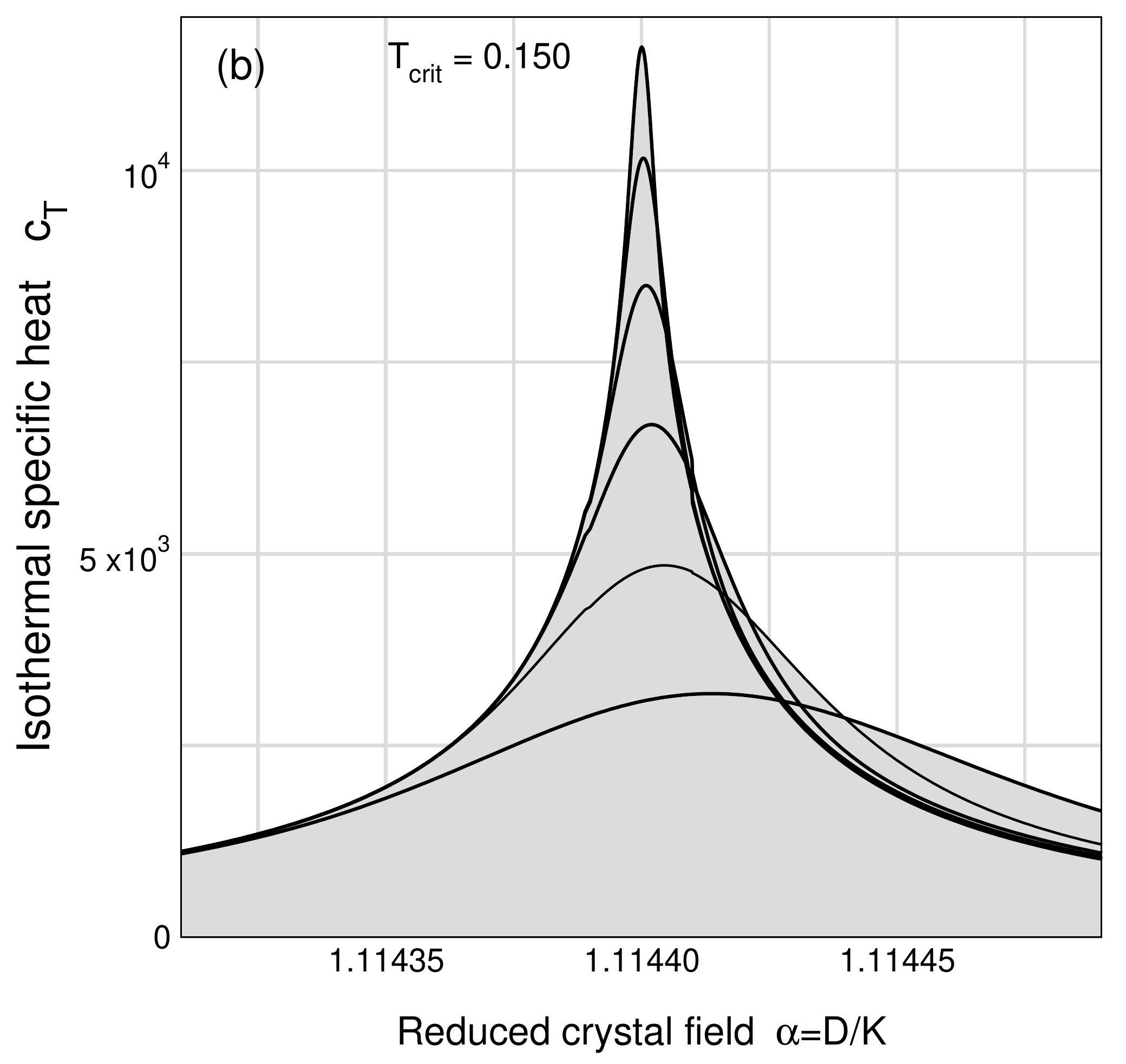}
\caption{Panel (a): Constant-field specific heat $c_\alpha$ versus the reduced temperature (units of K), for $\alpha_\textsf{crit} = 1.046$ (black circle), $\alpha_\textsf{crit} = 1.151$ (blue square) and $\alpha_\textsf{crit} = 1.246$ (red triangle) for the BC model on a $d = 2$ DHL with $n=10$ generations. Letters indicate the corresponding phase-transition points (first-order) on the frontiers of the phase diagram shown in Figure \ref{Fig2} (left panel). Arrows indicate the corresponding continuous phase-transition peaks according to the color labels. Panel (b): Isothermal specific heat at $T_\textsf{crit}=0.150$ for DHL's of sizes $L=2^n$, $n=10$ to $15$ from bottom to top. }
\label{Fig9}
\end{center}
\end{figure}

The specific heat $c_\alpha$  can be straightforwardly obtained by numerical differentiation of plots  like the ones shown in Figure \ref{Fig8}. For a general view, we explore the specific heat at the OP-F and F-P transitions at the reentrant regions (delimited by the dashed lines in Figure \ref{Fig2}) choosing three fixed-values of $\alpha \in (\alpha_\textrm D,  \alpha_\textrm B)$ and analyze the behavior of $c_{\alpha}$ when crossing the phase transition boundary at points below (point D), at (point C) and above TCP in the diagrams shown in Figure \ref{Fig2}. The left panel of Figure \ref{Fig9} displays the specific-heat plots for these three values of $\alpha_\textsf{crit}$ in this range, for the model defined on a $d=2$ DHL with $n=10$ generations. The plots for critical $\alpha_\textsf{crit}= 1.046$ and $1.151$ are drawn crossing the frontier OP-F at the first-order transition segment at low temperatures, while the one for $\alpha_\textsf{crit}=1.248$ is drawn crossing the frontier at the continuous transition segment. In all cases, the plots present two maxima one associated with the OP-F transitions and the other with the F-P one, both occurring in the reentrant region of the phase diagram. The   peaks indicated by arrows correspond to continuous transitions, while the others, points $D$ and $C$ at very low temperatures (below $T_\textsf{TCP}$) correspond to a first-order phase transition. 
For graphic scale reasons, the plots are shown for the BC model defined in a DHL with a low number of hierarchies ($n=10$). In general, the peak value of the specific heat increases, and its position shifts as the size of the lattice  increases.

Following again Zierenberg \emph{et al} \cite{zierenberg15} we now consider the fixed-temperature field-derivative of the spin-spin interaction $u$, which the appropriated partial amount of energy that captures its singular behavior as pointed out in \cite{zierenberg17}. Hence, we analyze such \emph{isothermal} specific-heat-like quantity 
\begin{equation}
\label{eq11}
c_{T}(\alpha)=\frac{\partial u}{\partial \alpha}\Big|_T,
\end{equation}
referring to the numerical derivative of $u$ as a function of $\alpha$, for a fixed temperature and varying the lattice size. At the transition the graph of $c_{T}(\alpha)$ displays a maximum that grows and becomes narrower indefinitely with increasing the lattice size $L=2^n$, $n$ being the number of generations, and whose position (called a pseudo-critical field) shifts as $n$ increases until reaching its genuine $\alpha$-critical value, similar to what happened with quadrupolar susceptibility. The right panel of Figure \ref{Fig9} shows the lattice-size $L=2^n$ dependence of $c_{T}(\alpha)$ for $n=10$ to $15$, for the model on a $d=2$ DHL at $T_\textsf{crit}=0.150$.
In general, the order of the transition, in this case, can also be manifested by the scaling behavior of the value of the maximum of the specific-heat-like quantity  $c_T^{[\max]}(L)$ for finite systems \cite{zierenberg17,janke03}.  The quantity $c_T^{[\max]}(L)$ is expected to exhibit the following   scaling behaviors: 
\begin{description}
    \item[(a)] $T_\textsf{crit}<T_\textsf{TCP}$: power-law behavior  $c_T^{[\max]} = a\;L^{p}$.
    \item[(b)] $T_\textsf{crit} > T_\textsf{TCP}$: logarithmic behavior like $c_T^{[\max]} = b + b' \ln L$.
    \item[(c)]  $T_\textsf{crit}=T_\textsf{TCP}$: both power-law and logarithmic behaviors are accepted but within distinct scaling intervals.
\end{description}

\begin{figure}
\begin{center}
\includegraphics[width=8cm]{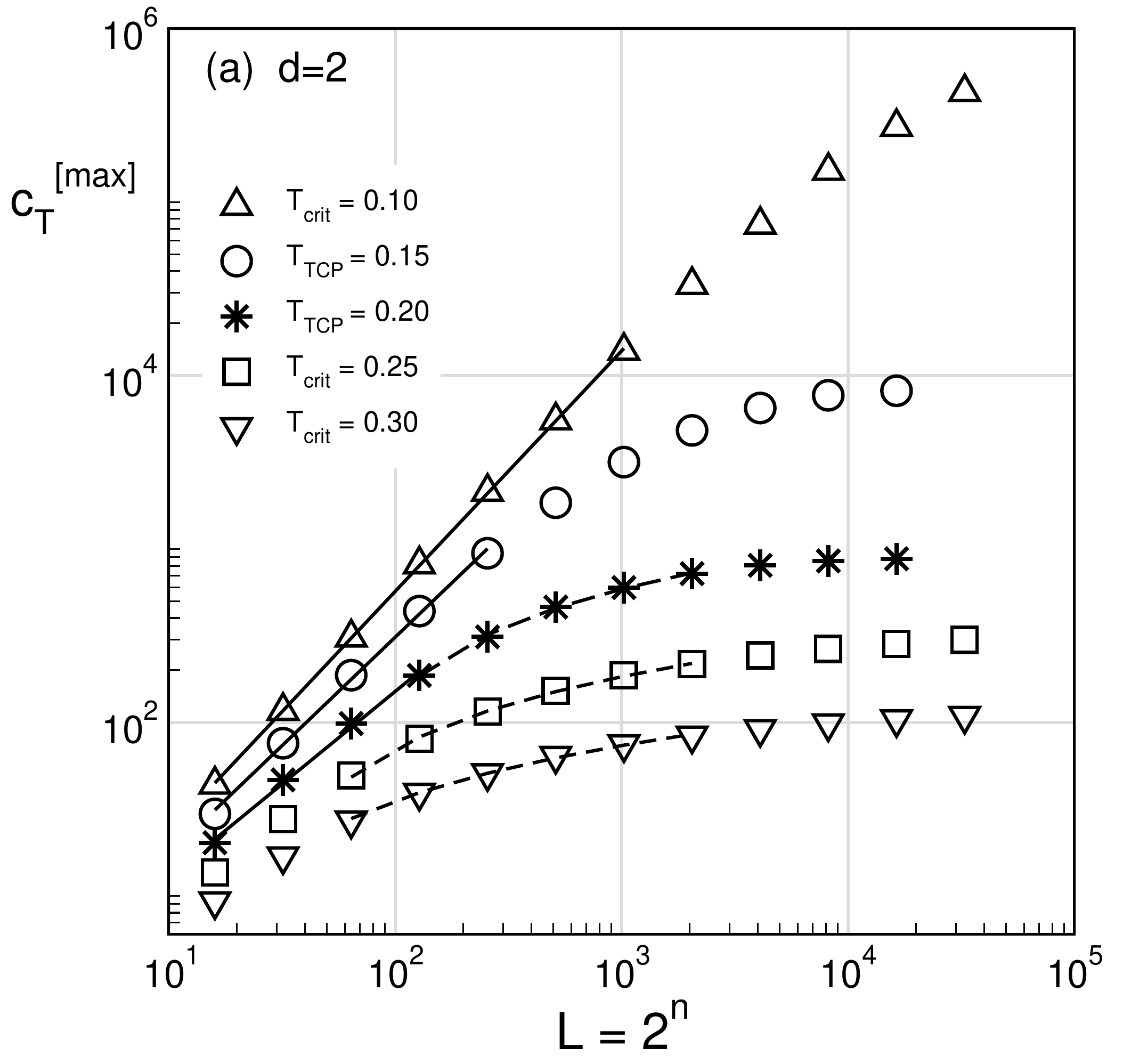}
\includegraphics[width=8cm]{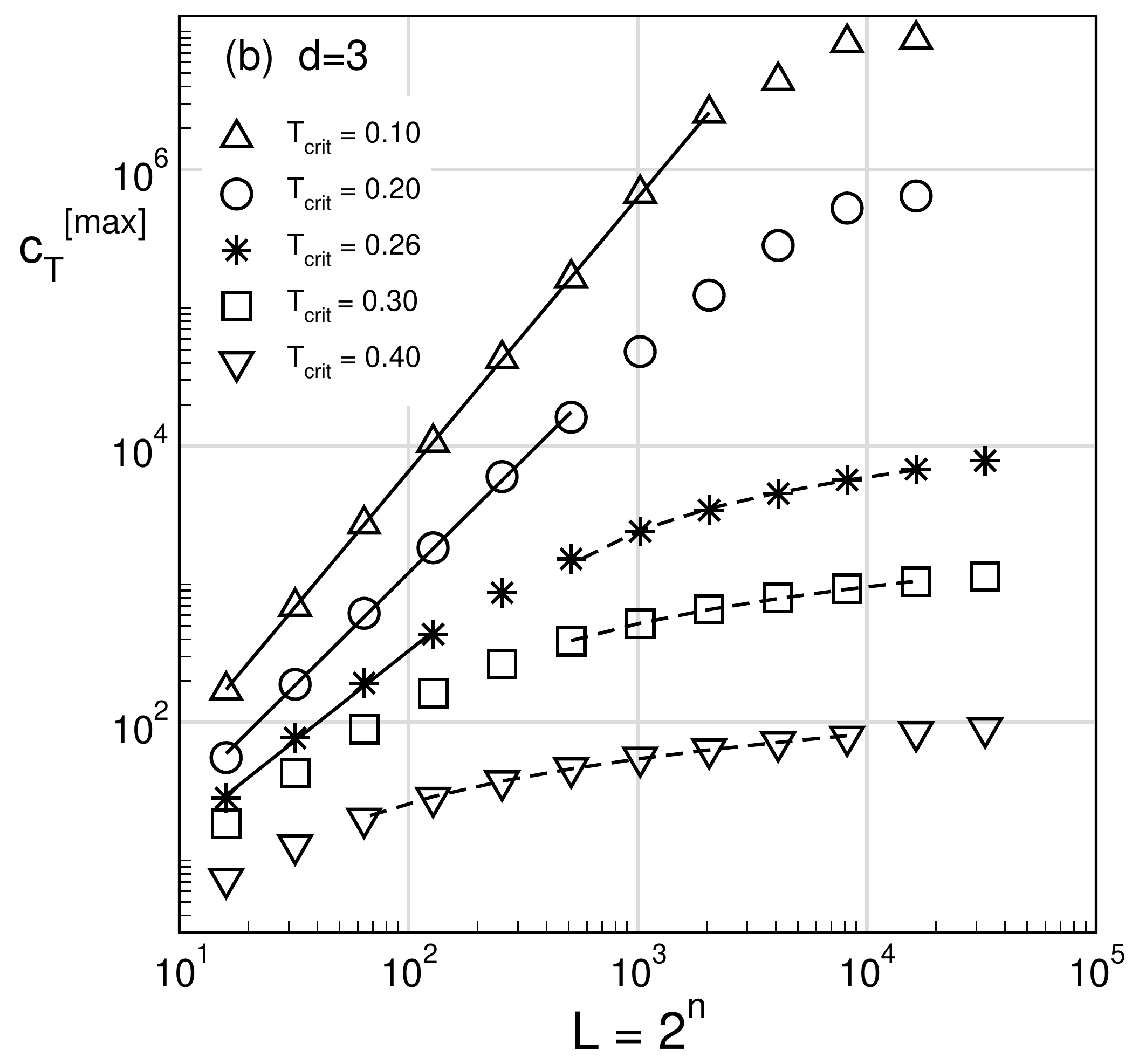}
\caption{Specific heat maximum $c_T^{\max}(L)$ versus DHL-size $L$ (log-log scales) for values of the critical temperature above, below, and at the estimated tricritical point (TCP). Panel (a) display plots for $T_\textsf{crit}=0.10 $ (up triangles), $0.15$ (circles), $0.200$ (stars TCP), and $0.20$ (stars TCP), $0.25$ (squares) and $0.30$ (down triangles) for a $d=2$ DHL.  Panels (b) exhibit the plots for $T_\textsf{crit}= 0.10$ (up triangles), $T_\textsf{crit}= 0.20$ (circles), $T_\textsf{TCP}= 0.26$ (stars), $0.30$ (squares), and $T=0.40$ (down triangles) for a $d=3$ DHL. In both panels straight lines indicate the interval with corresponding power-law fitting curves $c_T^{\max}(L)\sim L^{x_d}$, and  dashed curves represent the interval of the logarithmic adjustments $c_T^{\max}(L)\sim \ln(L)$.}
\label{Fig10}
\end{center}
\end{figure}

Figure \ref{Fig10} (a) illustrates five graphs of the $c_T^{[\max]}(L)$ with power-law scaling behavior for the cases where discontinuous transitions (first order) occur in $T_\textsf{crit} = 0.10$, $0.15$ and $0.20$, and with logarithmic scaling for continuous transitions for $T_\textsf{TCP} =0.20$, $T_\textsf{crit} =0.25$ and $0.30$, for the model BC defined in DHL with $d = 2$. Figure \ref{Fig10}(b) illustrates $c_T^{[\max]}(L)$ similar behavior for $d=3$-DHL but for the temperatures $T_\textsf{crit} = 0.10$ and $0.20$, $T_\textsf{TCP}=0.26$ and $T_\textsf{crit} = 0.30$ and $0.40$ cases. We note that the size range where the good fit occurs decreases as the temperature approaches $T_\textsf{TCP}$ either above or below. At temperatures close to the estimated $T_\textsf{TCP}$ both types of adjustments are observed, however, in distinct size ranges. At low values of $L$ the power-law type adjustment fits better, while for larger lattices the logarithmic type adjustment prevails. 
\newpage
Table \ref{Tab3} displays the coefficients $a$, $p$, $b$ and $b'$, the corresponding $R$ squared parameters and the scaling intervals for the  fitting curves of $c_T^{[\max]}(L)$ for the plots shown in Figure \ref{Fig10}(a) and (b).

\begin{table}[ht]
\renewcommand{\arraystretch}{1.2}
\setlength{\tabcolsep}{4pt}
  \centering 
\begin{tabular}{|c |c|c c|c|c|c|l|}
\hline
d & $T_\textsf{crit}$ &$a$ & $p$ &  $b$ & $b'$ & $R^2$ parameter & scaling interval  \\ \hline
&$0.10$ & $0.96(4)$ & $1.387(6)$  &  $\cdots$ & $\cdots$ & 0.999976 & $n=4 \to 11$\\  
&$0.15$ & $1.0(12)$ & $1.25(3)$  &  $\cdots$ &  $\cdots$ & 0.999252 &$n=4 \to 9$ \\  
2 &$0.20$ & $1.1(2)$ & $1.07(4)$  & $\cdots$ &  $\cdots$ & 0.998249 &$n=4 \to 7$ \\ 
&$0.20$ & $\cdots$ & $\cdots$ & $ - 764(25)$  & $196(6)$ & $0.999402$ & $n=7 \to 11$ \\  
&$0.25$ &$\cdots$ & $\cdots$ & $ - 156(4)$  & $49.2(7)$  & $0.999643$ & $n=6 \to 11$  \\  
&$0.30$ &$\cdots$ & $\cdots$ &$ -41(3) $  & $16.6(4)$ & $0.998581$ & $n=6 \to 11 $ \\  \hline
&$0.10$ & $0.70(18)$ & $1.982(3)$  &  $\cdots$ & $\cdots$ & 0.999991 & $n=4 \to 10$\\  
&$0.20$ & $0.60(11)$ & $1.64(2)$  &  $\cdots$ & $\cdots$ & 0.999559 & $n=4 \to 10$\\  
3 &$0.26$ & $0.80(16)$ & $1.31(13)$  &  $\cdots$ & $\cdots$ & 0.998942 & $n=4 \to 7$\\
&$0.26$ & $\cdots$ & $\cdots$  & $-8200(300)$ & $1540(30)$  &   0.999081& $n=9 \to 14$\\  
&$0.30$ & $\cdots$ & $\cdots$  & $-800(20)$ & $191(3)$  &   0.999462& $n=9 \to 14$\\  
&$0.40$ & $\cdots$ & $\cdots$  & $-32.2(8)$ & $12.6(1)$  &   0.999758& $n=6 \to 13$\\ \hline
\end{tabular}
  \caption{Adjustment coefficients $a$ and $p$ for the power-law fitting $c_T^{[\max]}(L) = a L^p$, $b$ and $b'$ for the logarithmic fitting $c_{\max}(L) = b + b'\ln L $ and the associated correlation coefficients for the plots shown in Figures \ref{Fig10}(a) and (b) for DHL-s with $d=2$ and $3$, respectively. The last column indicates the generation interval for the size scaling region $L=2^n$. Parentheses indicate the standard deviation.}
    \label{Tab3}
\end{table}

\section{Entropy and Enthalpy}\label{sec5}
\begin{figure}[ht]
\begin{center}
\includegraphics[width=8cm]{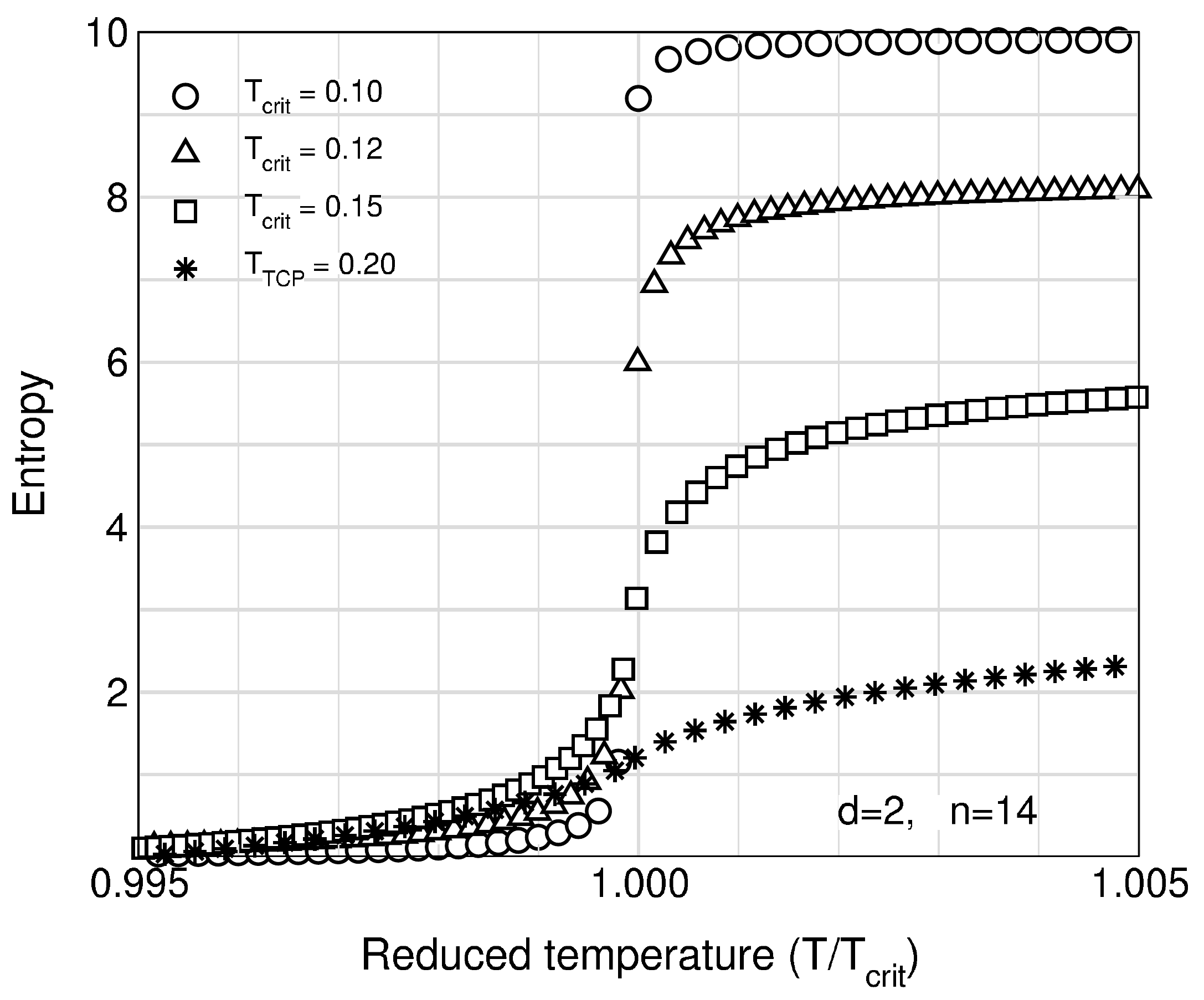}
\includegraphics[width=8cm]{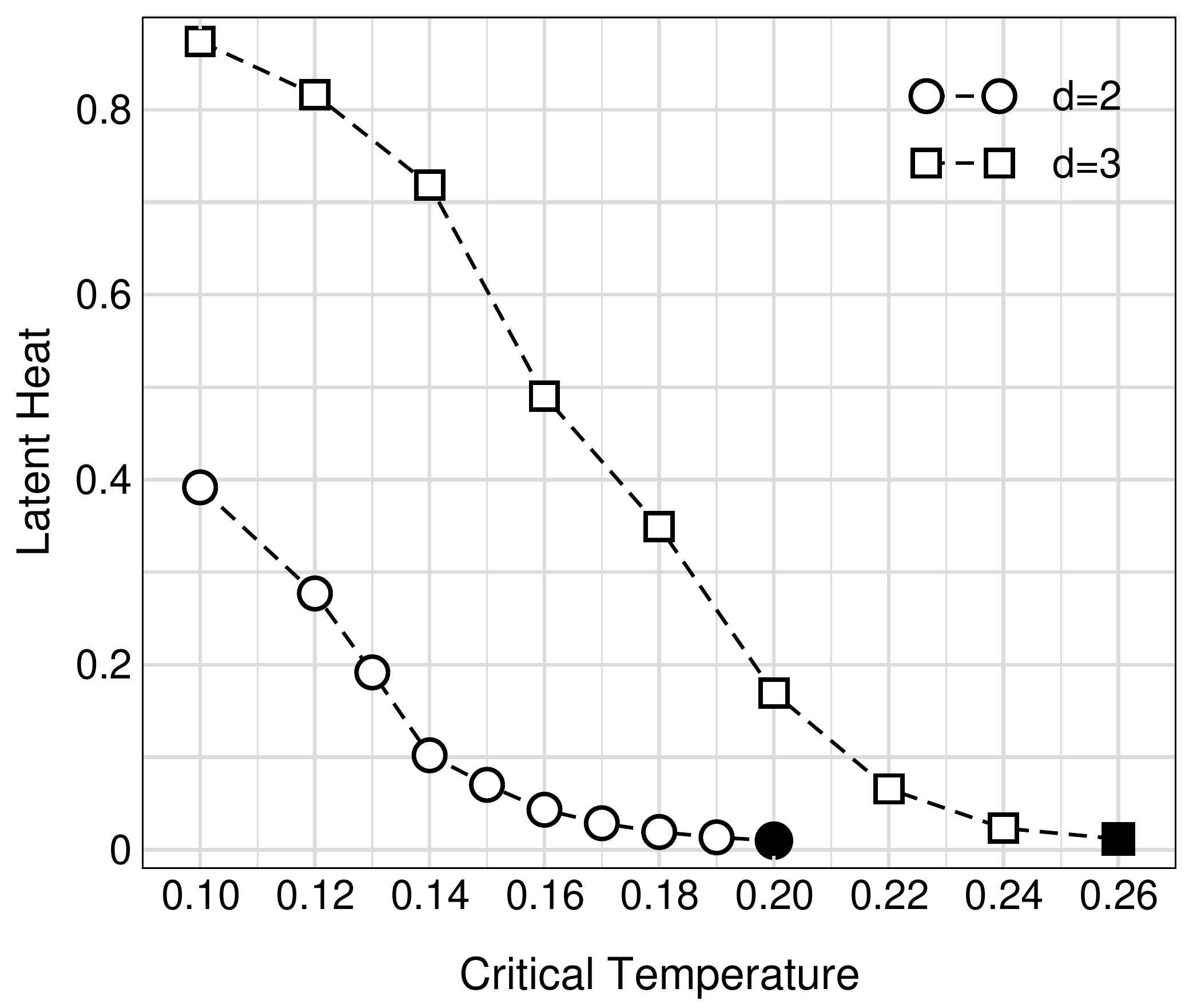}
\caption{Left panel: Entropy density $s(T)$ in units of $J$ as a function of $T/T_c$ on the F-OP transition frontier for the BC model in $d = 2$-DHL. Legends indicate the corresponding critical temperature values. Right panel: Latent heat as a function of the temperature on the critical transition line of the first-order phase evaluated for $L^{14}-$DHL size.  Legends indicate the DHL dimension and solid black symbols mark the corresponding estimated tricritical points.}
\label{Fig11}
\end{center}
\end{figure}

The present methodology also enables us to calculate entropy density $s$ from  the specific heat $c_\alpha$ given by (\ref{eq9}). In this case, the entropy density is obtained from the numerical integration of the constant crystal-field specific heat divided by the temperature, i.e.
\begin{equation}
\Big(\frac{\partial s}{\partial T}\Big)_\alpha =  \frac{c_\alpha}{T}, \qquad  c_\alpha \equiv \Big(\frac{\partial u}{\partial T}\Big)_\alpha \label{eq12}
 \end{equation}
For temperatures along the frontier of a second-order phase transition, there is a gradual change in the $s(T)$ concavity sign in $T_c$. However, for temperatures below TCP along the first-order transition frontier segment one observe an abrupt variation in the entropy density $s(T)$.
Figure \ref{Fig11} (left panel) shows the entropy density $s(T) $ as a function of the reduced temperature for a few critical temperature values below and at the tricritical point, as indicated in the legend for a $d=2$ DHL with $n=14$ generations. Similar plots (not shown) display the same patterns for a $d=3$ DHL. 
As the transition temperature decreases along the boundary, the discontinuity becomes more accentuated, demonstrating that the present methodology allows access to  the region where first-order transition occurs in the phase diagram, generally not   accessed by other methods, but recently viewed and explored via the transfer-matrix method \cite{jung17}.

To check the locus of the tricritical point, we directly calculate the enthalpy density values $\Delta h = T \Delta s $,    along the first-order transition line, in this case, called the \emph{latent heat} of the transition. According to phase-transition theory, the latent heat must be non-zero for the first-order transitions, whereas for a second-order transition we have no latent heat. In right panel of Figure \ref{Fig11}, we present the temperature latent heat dependence for DHL's with $n=14$ generations and $d=2$ and $d=3$ dimensions. The plots indicate that the location of the tricritical point is at $T_\textsf{TCP} \simeq 0.20$ for $d = 2$ and $T_\textsf{TCP} \simeq 0.26$ for $ d = 3 $ recovering the values obtained from the FSS analysis of isothermal susceptibility and specific heat discussed above. To obtain these results, we use temperature increments of the order of $10^{-6}$ in the calculation of the integration of the numerical entropy and estimate the values of $\Delta s$ for the points immediately before and after the corresponding transition point. More accurate calculations on larger lattices can be implemented with the help of more powerful computational resources. However, within our numerical limitations, we found that for a fixed temperature, the values of $\Delta s$ decrease as the lattice size increases. However, the convergence for $\Delta s \to 0$ at the tricritical point occurs at the same temperature value, respecting the range of uncertainties of the numerical calculation. These results are also compatible with the multifractal properties of the local magnetization described in \cite{rochaneto18} and signaled here by the thermodynamic behavior of the model. 

\section{Summary of Results and Conclusions}\label{sec6}
This work expands, deepens, and complements the study of the Blume-Capel model carried out  by the authors in the reference \cite{rochaneto18}.  Here we investigate the  thermodynamical behavior of the  BC model in the re-entrant region of the phase diagram focusing on the boundary between the ferromagnetic and \emph{ordered paramagnetic} phases (segment B-C-D-E of Figure \ref{Fig2}). In particular, we explore in detail the behavior of some relevant thermodynamical quantities along the critical boundary at the vicinity of the tricritical point. We consider the BC model defined on the diamond hierarchical lattice with dimensions $d=2$ and $3$,   and we   obtain the \emph{exact} numerical local values for the internal energy and the order parameters  densities (magnetization $m$ and quadrupole moment $q$). We then numerically explored $n_0=1-q$ the relevant density of sites in the spin state $S=0$ and its response function, the isothermal quadrupolar susceptibility $\chi_T$ in Section \ref{sec3}.  Furthermore, we studied the average exchange energy density $u$ and its response functions, the constant-field $c_\alpha$ and isothermal $c_T$ specific heats, in Section \ref{sec4}. Finally, we analysed the entropy density $s$ and the latent heat associated with the first-order transition  in Section \ref{sec5}. \\
    
Below we summarize the main results and conclusions:
\begin{itemize}
\item The isothermal quadrupolar susceptibility $\chi_T(\alpha)$ was investigated in the reentrant F-OP frontier of the phase diagram at fixed temperatures below, above, and at the TCP for $d=2$ and $d=3$ DHLs with high sizes. We obtained very narrow and intense peaks at temperatures below $T_\textsf{TCP}$ but rather smaller ones for $\alpha \geq \alpha_\textsf{crit}$, which can be perceived by the shape of the plots shown in the panels of Figure \ref{Fig6} in normalized scales.  At $T \simeq T_\textsf{TCP}$, we observe narrow and intense peaks but with no significant discontinuity between values around $\alpha_\textsf{crit}$, as expected for the onset of a continuous transition. The plots for the model on a 3-d DHL exhibit a more prominent asymmetric pattern.   Furthermore, a detailed study of the variation of its maximum value $\chi^{[\max]}_T(L)$ as a function of lattice size $L=2^n$, using finite-size scaling techniques, shows that $\chi^{[\max]}_T$ grows following a power-law with first-order corrections at temperatures below $T_\textsf{TCP}$ (first-order transition) and behaves logarithmically at the boundary where continuous transitions occur. Such behaviors were verified in the BC model defined in both the DHLs of dimensions $d=2$ and $d=3$.
\item The constant-field $c_\alpha(T)$ and \emph{isothermal} $c_T(\alpha)$ specific heat were calculated and explored from the numerical derivatives of the internal exchange energy per bond $u$, associated with the exchange coupling between the spins. This response function captures the singular behavior in the change in sign of the $u$-concavity at temperatures below $T_\textsf{TCP}$, as shown in Figure 8. Above this value, the $u$-concavity variation becomes smooth as temperatures are raised to higher values. A detailed finite-size scaling analysis of $c_T^{[\max]}$ reveals that at temperatures below $T_\textsf{TCP}$ the value of $c_T^{[\max]}$  grows following a power-law behavior (in zero order) over a wide range from smaller lattice sizes. However, at the boundary where the continuous transitions occur, the growth of $c_T^{[\max]}$ is of the logarithmic type, occurring however in intervals of larger lattice sizes. Moreover, at temperatures around $T_\textsf{TCP}$, both behaviors are observed, in smaller but distinct scaling intervals, both with expressive values for the control parameter $R^2\simeq 1$ as displayed in Figures \ref{Fig10}(a) and (b). The corresponding coefficients, exponents, scaling-intervals and $R^2$ parameter of the scaling-functions for both BC model on the $d=2$ and $d=3$ DHLs are exhibited in Table \ref{Tab3}.
\item Entropy density and latent heat: the entropy density $s(T,\alpha)$ which was obtained from the numerical integration of the specific heat divided by the temperature $c_\alpha(T)/T$, reveals the singular behavior in its concavity. The gaps observed at critical temperatures over the transition boundary in the region below the tricritical point make it possible to obtain the \emph{latent heat} associated with the first-order transition. The decreasing up to zero of this quantity with the increase of the temperature on the transition line made it possible to accurately retrieve the tricritical-point temperatures for the BC model in the DHLs with dimensions $d=2$ and $3$. 
\end{itemize}

The results described above demonstrate that the present approach is quite adequate to scrutinize the thermodynamic behavior of the BC model in the vicinity of the tricritical point. The response functions associated with the magnetization, quadrupole moment, and entropy densities exhibit distinct and characteristic behaviors when $T >T_\textsf{TCP}$ (second-order transition) or $T < T_\textsf{TCP}$ (first-order transition). In particular, they consistently unveil the characteristic discontinuities of first-order transitions below $T_\textsf{TCP}$. Moreover, the scaling invariance property of the DHLs makes it possible to perform finite-size scaling analysis on intervals comprising large-sized systems allowing us to explore the critical behavior of pure BC model at its tricritical point. A careful study of the critical exponents associated with quadrupolar order-parameter $q$ of the BC model, at the tricritical point, however, requires a more powerful and sophisticated numerical analysis that we hope to carry on in the near future.

Finally, the thermodynamic properties of the BC model can also be explored in other regions of the parameter space $(T, \alpha)$. In particular, the characterization of the \emph{ordered-paramagnetic} phase at low temperatures and at values of $\alpha \gtrsim \alpha^\star$  impel our current interest. The generalization of the pure BC model to consider competitive disorder (or dilution) in the exchange couplings and/or the crystal-field strength also motivate our coming works.  
\section*{Declaration of competing interest}
The authors declare that they have no known competing financial interests or personal relationships that could have
appeared to influence the work reported in this paper.
\section*{Acknowledements}
M. J. G. Rocha-Neto thanks CNPq (Brazilian agency for research support) for the scholarship received during his doctoral program (grant 141185/2013), during which most of the results presented here were obtained.

\end{document}